\documentclass[a4paper,fleqn,usenatbib]{mnras}

\usepackage[T1]{fontenc}
\usepackage{ae,aecompl}

\usepackage{graphicx}	
\usepackage{amsmath}	
\usepackage{amssymb}	

\title[Proper Motion of Draco Dwarf Galaxy]{Proper Motion of the Draco Dwarf Galaxy from Subaru Suprime-Cam Data}

\author[D. I. Casetti-Dinescu \& T. M. Girard]{
Dana I. Casetti-Dinescu$^{1,2}$\thanks{E-mail: danacasetti@gmail.com}
and Terrence M. Girard$^{2}$\thanks{Present address: 14 Dunn Rd, Hamden, CT 06518, USA}
\\
$^{1}$Physics Department, Southern Connecticut State University, 
New Haven, CT 06515-1355, USA \\
$^{2}$Astronomy Department, Yale University, P.O. Box 208101,
New Haven, CT 06520-8101, USA \\
}

\date{Accepted XXX. Received YYY; in original form ZZZ}

\pubyear{2016}

\begin{document}
\label{firstpage}
\pagerange{\pageref{firstpage}--\pageref{lastpage}}
\maketitle

\begin{abstract}
We have measured the absolute proper motion of the Draco dwarf spheroidal galaxy using Subaru Suprime-Cam images taken
at three epochs, with time baselines of 4.4 and 7 years.
The magnitude limit of the proper-motion study is $i=25$, thus allowing for
thousands of background galaxies and Draco stars to be used to perform extensive
astrometric tests and to derive the correction to an inertial reference frame.
 The derived proper motion is 
$(\mu_{\alpha}, \mu_{\delta}) =(-0.284\pm0.047, -0.289\pm0.041)$ mas yr$^{-1}$.
This motion implies an orbit that
takes Draco to a pericenter of $\sim 20$ kpc;
a somewhat disruptive orbit suggesting that tides 
might account for the rising velocity-dispersion profile of Draco seen
in line-of-sight velocity studies. 
The orbit is only marginally consistent with Draco's membership to the
vast polar structure of Galactic satellites, in contrast to a
recent HST proper-motion measurement that finds alignment very likely.
Our study is a test case to
demonstrate that deep imaging with mosaic cameras of appropriate resolution
can be used for high-accuracy, ground-based proper-motion measurement.
As a useful by-product of the study, we also identify two faint brown-dwarf candidates in
the foreground field.
\end{abstract}





\section{Introduction}

The prevalence of structure in the halo of the Milky Way (MW), with orbital debris reaching out to its farthest
extent, is a relatively recent discovery that gives visible testament to the assembly history of our Galaxy.
Our knowledge of this structure is due primarily to large-area/all-sky, multiband 
photometric surveys such as SDSS, and more recently PanSTARRS and the Dark Energy Survey (DES), to name a few. 
Adding to the traditional eleven dwarf spheroidal galaxy satellites of the MW known 
since the mid-twentieth century, tens of more have 
been discovered along with distant globular clusters and various tidal streams and features 
\citep[e.g.,][]{kop15,mar15}.
Ascertaining the properties of these star systems, their relation to our Galaxy and to one another, 
provides the most detailed and accurate picture of galaxy-formation on small scales where 
tensions with current cosmological models exist. 

One such property of these star systems that too often remains unmeasured is their tangential velocities, or 
absolute proper motions. These motions
are critical to deciphering the dynamical aspects of the assembly process of the MW. 
Because of the needed astrometric precision, most proper-motion studies of distant satellites have made use of
the Hubble Space Telescope ($HST$) 
\citep[see][review]{vdm15}, and much is 
expected in the not-too-distant future from the ongoing ESA space mission, Gaia.
Here we introduce a first result from a new, ground-based program that instead uses archival images taken with 
wide-field imagers with appropriate scale ($\sim 0\farcs2$/pixel) mated to
large (4 to 8m class) telescopes. 
Archives now include images separated by as much as 10 years, that are sufficiently deep (to $V \sim 25$) and wide 
(tens of arcmin). 
Such images include tens of thousands of background galaxies that can be used to anchor the measured proper motions
to an inertial reference system.

There are a number of reasons to undertake such a program. 
First, it is complementary to $HST$ and Gaia. 
It is often the case that $HST$ will not have first-epoch observations appropriate to determine proper motions for the
newly discovered ultra faint dwarf galaxies and 
remote globular clusters, 
while Gaia's limiting magnitude of $V\sim 20$ will not reach the main sequence of these rather sparse systems.
Second, most existing and planned (e.g., LSST) wide-field, deep imaging surveys have yet to have their astrometric
potential properly exploited, in the context of providing large numbers of scientifically useful 
proper motions. 
Tapping this potential requires an astrometric modeling of the detector and telescope system, and here we lay the
groundwork with one such methodology. 
Third, the shear number of galaxies and stars per unit area and magnitude in these deep, wide-field images
is somewhat novel in this type of work (and represents a precursor to LSST). 
These numbers allow us to explore various sources of systematic errors,
and better understand the limitations of our proper-motion measurement. 
Fourth, there is an interesting by-product of this type of wide and deep
proper-motion study: finding fast-moving, faint objects as candidate cool dwarfs and white dwarfs. 

We present here a test-case 
study in the field of the Draco dwarf spheroidal galaxy (dSph) using images taken with the 
Subaru Prime Focus Camera, Suprime-Cam \citep{miy02}.
The data span three epochs with time baselines of 4.4 and  7 years, and
cover a total of 0.2\footnote{0.148 sq. degrees for the determination of the mean motion of Draco, see Section 4.} square degrees to a limiting magnitude of $i \sim 25.0$. 
The resulting proper-motion measurement for Draco indicates a rather eccentric orbit 
with a pericenter of $\sim 20$ kpc, and currently having just passed its apocenter. 
Within uncertainties, Draco's orbit is marginally
consistent with membership to the Vast Polar Structure described by \citet{paw13}, although it shows a 
poorer alignment with this structure than what the recent $HST$ proper-motion measurement indicates 
\citep[][hereafter P15]{pr15}.




\section{Observations}
We have searched the Subaru Mitaka Okayama Kiso Archive system (SMOKA), \citep{baba02} for observations taken with 
the Suprime-Cam \citep{miy02} centered on 
the Draco dwarf galaxy. We found data taken in 2001, 2004, 2005 and 2008. The 2004 data set was very shallow and of 
poor quality, therefore we have discarded it from further analysis. The data sets to be used 
in our proper-motion study (i.e., those with considerable overlap) 
total 94 individual exposures. The characteristics of these exposures -- date, filter, exposure time (in seconds), 
and number of frames per data set -- are listed in Table~\ref{tab:tab1}. Eventually, due to proper-motion determination limits,
we used only 85 exposures for the final proper motion catalog.
\begin{table}
\caption{Observation Log}
\label{tab:tab1}
\begin{tabular}{llll}
\hline
\multicolumn{1}{c}{Date} & \multicolumn{1}{c}{Filter} & \multicolumn{1}{c}{Exp.} & \multicolumn{1}{c}{N} \\
\hline
\multicolumn{1}{l}{2001/03/19-20} & \multicolumn{1}{l}{$V$} & \multicolumn{1}{r}{10} & \multicolumn{1}{r}{8} \\
\multicolumn{1}{c}{''} & \multicolumn{1}{l}{''} & \multicolumn{1}{r}{200} & \multicolumn{1}{r}{1} \\
\multicolumn{1}{c}{''} & \multicolumn{1}{l}{''} & \multicolumn{1}{r}{500} & \multicolumn{1}{r}{2} \\
\multicolumn{1}{c}{''} & \multicolumn{1}{l}{''} & \multicolumn{1}{r}{600} & \multicolumn{1}{r}{3} \\
\multicolumn{1}{c}{''} & \multicolumn{1}{l}{''} & \multicolumn{1}{r}{650} & \multicolumn{1}{r}{2} \\
\multicolumn{1}{c}{''} & \multicolumn{1}{l}{''} & \multicolumn{1}{r}{800} & \multicolumn{1}{r}{1} \\
\multicolumn{1}{c}{''} & \multicolumn{1}{l}{$R_C$} & \multicolumn{1}{r}{10} & \multicolumn{1}{r}{5} \\
\multicolumn{1}{c}{''} & \multicolumn{1}{l}{''} & \multicolumn{1}{r}{200} & \multicolumn{1}{r}{4} \\
\multicolumn{1}{c}{''} & \multicolumn{1}{l}{''} & \multicolumn{1}{r}{300} & \multicolumn{1}{r}{4} \\
\multicolumn{1}{c}{''} & \multicolumn{1}{l}{''} & \multicolumn{1}{r}{360} & \multicolumn{1}{r}{4} \\
\multicolumn{1}{l}{2005/08/05} & \multicolumn{1}{l}{$V$} & \multicolumn{1}{r}{300} & \multicolumn{1}{r}{5} \\
\multicolumn{1}{c}{''} & \multicolumn{1}{l}{$i$} & \multicolumn{1}{r}{240} & \multicolumn{1}{r}{10} \\
\multicolumn{1}{l}{2008/04/02-04} & \multicolumn{1}{l}{V} & \multicolumn{1}{r}{10} & \multicolumn{1}{r}{9} \\
\multicolumn{1}{c}{''} & \multicolumn{1}{l}{''} & \multicolumn{1}{r}{120} & \multicolumn{1}{r}{10} \\
\multicolumn{1}{c}{''} & \multicolumn{1}{l}{$I_C$} & \multicolumn{1}{r}{10} & \multicolumn{1}{r}{6} \\
\multicolumn{1}{c}{''} & \multicolumn{1}{l}{''} & \multicolumn{1}{r}{200} & \multicolumn{1}{r}{20} \\
\hline
\end{tabular}
\end{table}

\begin{figure}
\includegraphics[width=\columnwidth,angle=-90]{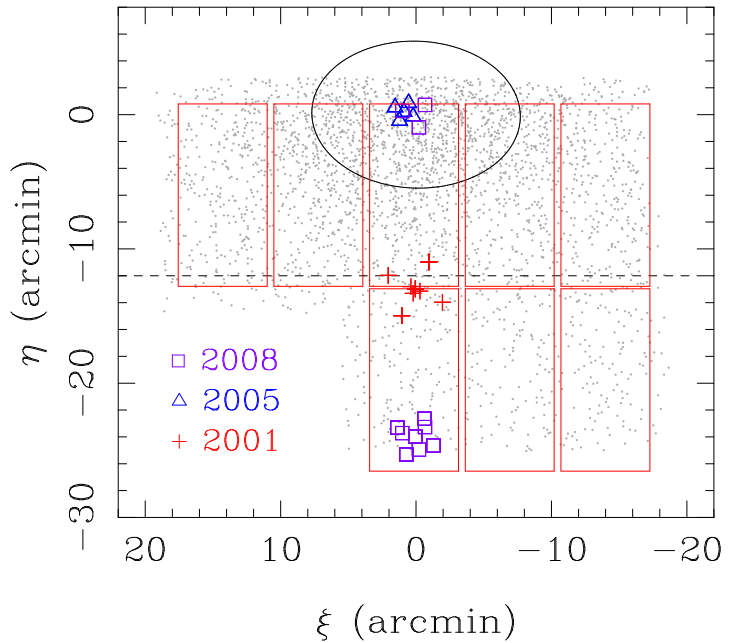}
\caption{
Spatial distribution of the data sets. Field centers of the Subaru exposures
are shown with large symbols, colour- and shape-coded by epoch of observation.
Gray points show every 20th object from our final proper-motion catalogue.
North is up, and East to the left. The origin of the plot is at Draco's 
center and the ellipse shows the extent of the core of the dwarf galaxy. 
In the final determination of the proper motion of Draco, we do not include
objects south of the dashed line, due to object classification constraints.}
\label{fig:fig1}
\end{figure}

In Figure~\ref{fig:fig1} we show the spatial coverage of the field of study in standard coordinates ($\xi,\eta$).
The origin is at Draco's center as adopted from \citet[][hereafter S07]{se07}.
The small gray dots represent objects from our final proper-motion catalogue (we plot every 20th object only);
the large symbols show the pointings of the Suprime-Cam exposures. 
Suprime-Cam is a mosaic of ten $2048\times4096$ CCDs located at the prime focus of the Subaru telescope
and covers a $34\arcmin\times27\arcmin$ field of view, with a pixel scale of $0\farcs20$ \citep{miy02}.
Our area of study would be slightly larger than the footprint of the Suprime-Cam, except for 
the missing data in the bottom, left corner. 
This is due to the 2001 data set; CCD w93c2 was not used and 
data from CCD w9c2 were astrometrically very poor (see Section 4). The ellipse indicates the extent of the core of 
Draco with semimajor axis, ellipticity and position angle taken from \citet{ode01}.


The archived data are extracted for each individual CCD, and, throughout the entire reduction process, we treat each
CCD as a single unit. Pre-processing, i.e., overscan, bias subtraction and flat fielding are performed using the 
packages SFRED1 for data before July 21, 2008
\citep{ya02,ou04}.
We have followed the steps listed in the Suprime-Cam data 
reduction manual\footnote{www.naoj.org/Observing/Instruments/SCam/sdfred} 
up to the distortion and differential geometric atmospheric dispersion corrections. 
We have {\bf not} included this step since it applies the corrections directly on the pixel data; we prefer to account for these effects in the astrometric solutions, as plate terms (see
Section 4.1). We have, however, used the PSF-size measurement step in SFRED1 that provides an average FWHM for 
each exposure and individual CCD, and this we use as input 
in the next step of detection and centering. The code SourceExtractor  \citep{be96}
is used to provide 
object detection, photometry, preliminary centers and object classification. 
Final object centers, in CCD pixel coordinates, are determined by fitting a two-dimensional elliptical Gaussian function
to each object's intensity data (Yale centering routines upgraded for CCD data from the original code described in \citet{leeva83}), 
with the centers from SExtractor serving as initial values for the nonlinear fitting algorithm.

Subaru's design of the prime focus corrector includes an atmospheric dispersion corrector (ADC) to correct for 
chromatic atmospheric dispersion \citep{miy02}. All the observations considered here had the ADC in use (we have verified
this by inspecting the correlation between the ADC position and the hour angle of the observation).

\begin{figure}
\includegraphics[width=\columnwidth,angle=0]{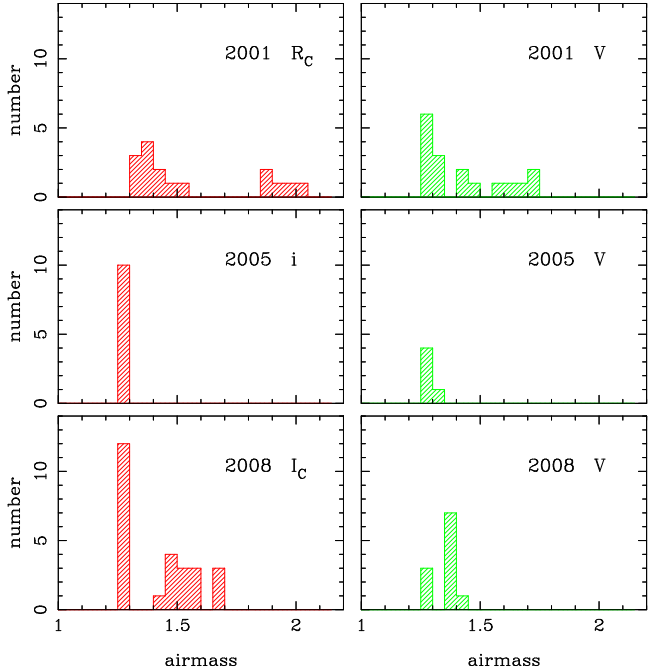}
\caption{Distribution in airmass, per filter and epoch.}
\label{fig:fig2}
\end{figure}

In Figure~\ref{fig:fig2} we show the airmass distribution for the three epochs, and for each filter.
The 2001 data have the largest airmass range, while the 2005 data have both the lowest average airmass and airmass range. 

\begin{figure}
\includegraphics[width=\columnwidth,angle=0]{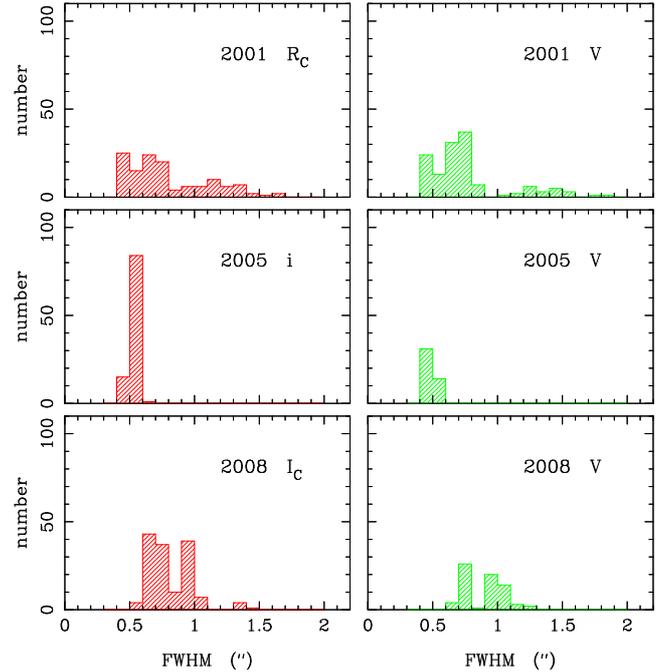}
\caption{Distribution in FWHM, per filter and epoch.}
\label{fig:fig3}
\end{figure}

In Figure~\ref{fig:fig3} we show the FWHM distribution for each epoch and filter. All Suprime-Cam chips are 
included in these distributions. The 2005 data have the best seeing. We will therefore use only these data for our object
classification.

\section{Photometry}
The Draco dwarf galaxy has an existing deep, large-area photometric dataset provided by S07 from 
observations taken with the MegaCam on the 3.6m CFHT. We use the $r$ and $i$ bands in the MegaCam system (in the AB system)
from this study to roughly calibrate our Subaru instrumental magnitudes. 
(Hereafter, when we use photometry from S07, we will label it as $i_{MegaCam}$ and $(r-i)_{MegaCam}$.)
Our calibrated magnitudes are only used for the purpose of selecting objects in broad ranges of $i$ magnitudes 
and $r-i$ colors to explore magnitude- and color-dependent systematics in the astrometry.
The calibration is done per individual chip and exposure
as an offset between instrumental and calibrated magnitude. 
Instrumental magnitudes are SExtractor isophotal magnitudes.  
For both instrumental $i$ and $I_C$ we have used $i_{MegaCam}$;
for instrumental Johnson $V$ we have used  $r_{MegaCam}$. 
Then, for each object, the calibrated magnitude is an average 
over the number of observations obtained with an iterative $3\sigma$ clipping.
The S07 data set covers the entire area of our proper-motion catalog, except for gaps between chips in the MegaCam 
mosaic detector; it is also shallower than ours in $i$ by $\sim 0.25$ mag. 
We note, however, that our faint limit is not set
by the detection threshold, but rather by the ability to center objects with the two-dimensional Gaussian fit
used in the astrometric analysis. 
Thus our centering requirement yields a set of objects that becomes incomplete at a signal-to-noise ratio of 
$\sim 30$ compared to the list of objects merely detected (e.g., at $S/N=10$, only $54\%$ of the
objects were centered with the two-dimensional Gaussian fit).
Thus the Subaru data are substantially deeper than the MegaCam data, and for this reason we give preference to
the Subaru object classification over that in S07. 

\section{Object Classification}
\begin{figure*}
\includegraphics[scale=0.8,angle=-90]{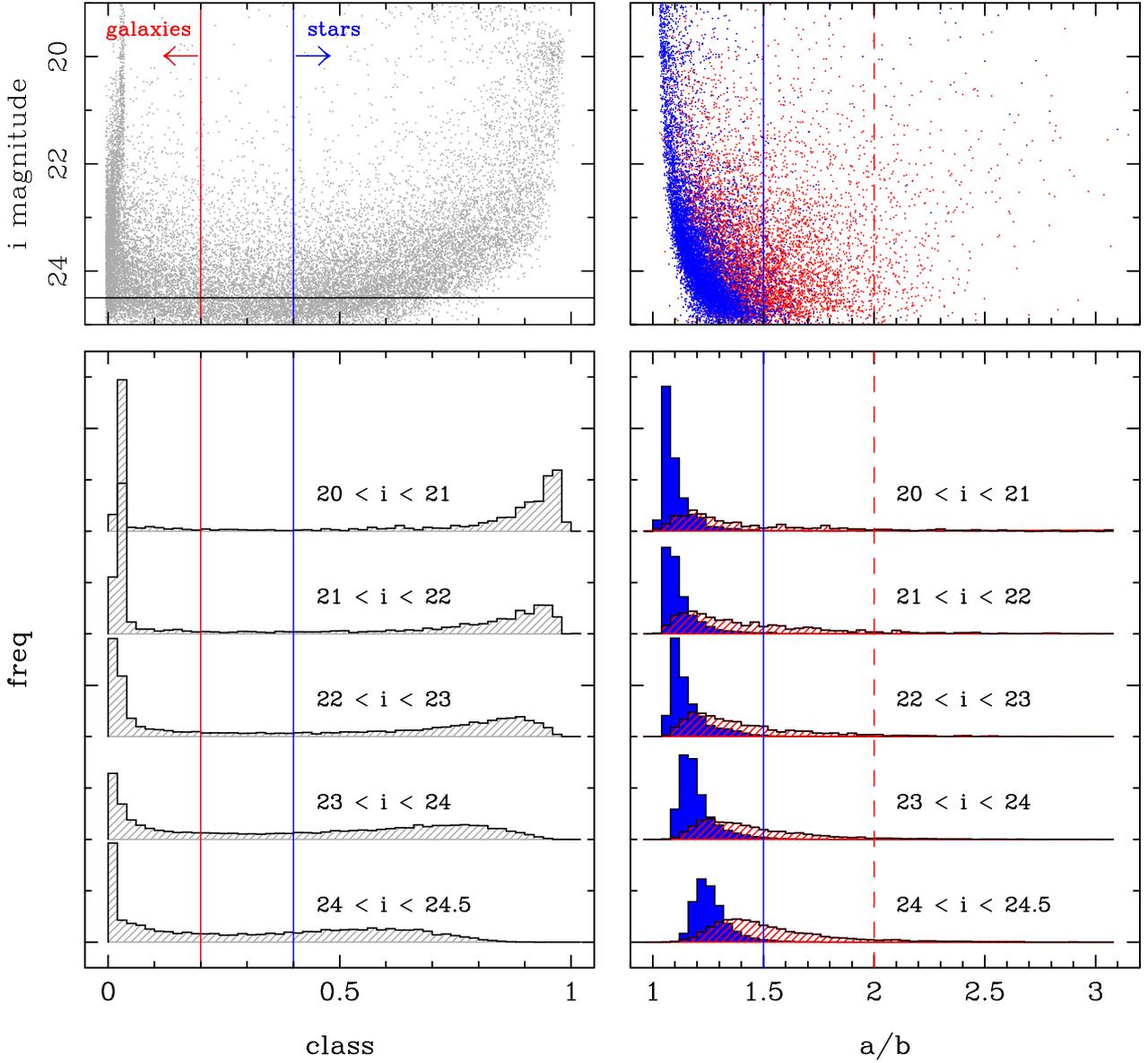}
\caption{Magnitude versus classification parameter (top left) and versus semimajor-to-semiminor ratio $a/b$ (top right).
First-pass selection for galaxies is those objects with 
class $\le 0.2$ (red line), and for stars, those objects with class $\ge 0.4$ (blue line).
Histograms of class and $a/b$ are shown for various magnitude bins in the corresponding lower panels.
First-pass selected galaxies (red) and stars (blue) are shown in the right panels.
The second-pass selection chooses only those galaxies that have $a/b \le 2.0$ (dashed line) and only 
those stars with $a/b \le 1.5$ (solid line). Fainter than $i=24.5$, the separation is no longer reliable.
}
\label{fig:fig4}
\end{figure*}

We employ the object classifier from SExtractor, which is a neural-network-based classifier that uses as input the FWHM.
For each object we determine the average (over the number of measurements) of the class parameter and of the 
ratio of the semimajor to semiminor axis, $a/b$, also given by SExtractor. 
We will use the object classification determined from only the 2005 data, which have a median seeing of $0\farcs54$.
This will restrict the area to be used in the final proper motion determination
to 0.148 square degrees, north of the dashed line in Figure~\ref{fig:fig1}.
In Figure~\ref{fig:fig4} we show $i$ magnitude as a function of average object parameter: 
class (top left panel) and $a/b$ (top right panel). Histograms of the class (left)  and $a/b$ (right) for
various magnitude bins are shown in the subsequent panels.
In a first-selection pass, we choose as stars objects with class $\ge 0.4$, 
and as galaxies objects with class $\le$ 0.2. This preliminary separation is highlighted
in all of the right panels: full histogram shows the $a/b$ for stars, and hashed
histogram for galaxies.
In the second-selection pass, we choose as galaxies to be used in 
our analysis, those with $a/b \le 2.0$; and as stars, those with $a/b \le 1.5$. 
Finally, we enforce a magnitude limit of
$i \le 24.5$, beyond which the star/galaxy separation is no longer reliable.

The class parameter histograms are bimodal in all magnitude bins, as expected, although the stellar peak
migrates to smaller values at fainter magnitudes.  This raises the issue of stellar contamination
of the galaxies, especially at the faint end.
We have attempted to make conservative cuts to minimize misclassification, leaving behind a substantial region of
unclassified objects.  Clean galaxy and star samples are the goal, rather than complete samples.

To assess the contamination of our galaxy sample by stars, due to low S/N, we perform the following test:
We construct new star and galaxy samples based on the entire data set, i.e., the 2001, 2005, and 2008 epoch data, using the same
class and $a/b$ cuts as above.
The median FWHM for these data is $0\farcs84$, or an increase of $55\%$ with respect to the 2005 data.
In Table~\ref{tab:tab2} we present the resulting numbers ($N$) of galaxies and stars, by magnitude bin, for the $0\farcs54$ and 
$0\farcs84$ data. We also give these as fractions ($f$) of the total number of objects considered 
(within the proper-motion catalog and same spatial area).
For the galaxies, the numbers do not differ significantly between the two samples based on differing FWHM observations,
until the last magnitude bin, where the sample with larger FWHM appears to show contamination. We will
see in the proper-motion determination, that this last bin is biased by the stellar field.
For the stars, numbers are larger in the better FWHM samples for the faintest 4 magnitude bins, indicating
that poorer FWHM moves stars into the unclassified domain.
Based on this, for galaxies, we will adopt $i=24.5$ as a faint limit for reliable classification.
For the stellar sample, we will limit the analysis to $i=24$.

\begin{table*}
\caption{Object-Classification Statistics at two values of the FWHM}
\label{tab:tab2}
\begin{tabular}{rrrrrrrrrr}
\hline
 & \multicolumn{4}{c}{Galaxies} & & \multicolumn{4}{c}{Stars} \\
\hline
 & \multicolumn{2}{c}{$0\farcs54$} & \multicolumn{2}{c}{$0\farcs84$} & & \multicolumn{2}{c}{$0\farcs54$} & \multicolumn{2}{c}{$0\farcs84$} \\
\hline
$i$ range & \multicolumn{1}{c}{$N$} & \multicolumn{1}{c}{$f$} & \multicolumn{1}{c}{$N$} & \multicolumn{1}{c}{$f$} & & \multicolumn{1}{c}{$N$} & \multicolumn{1}{c}{$f$} & \multicolumn{1}{c}{$N$} & \multicolumn{1}{c}{$f$} \\
\hline
19-21 & 670 & 0.266 & 674 & 0.268 & & 1650 & 0.655 & 1694 & 0.672 \\
21-22 & 1165 & 0.447 & 1145 & 0.439 & & 1178 & 0.452 & 1207 & 0.463 \\
22-23 & 2258 & 0.439 & 2535 & 0.435 & & 2645 & 0.454 & 2556 & 0.438 \\
23-24 & 5568 & 0.376 & 5474 & 0.370 & & 7334 & 0.496 & 6686 & 0.452 \\
24-24.5 & 4156 & 0.347 & 4127 & 0.345 & & 5503 & 0.460 & 4800 & 0.401 \\
24.5-25 & 3389 & 0.302 & 3481 & 0.310 & & 4172 & 0.372 & 3157 & 0.283 \\
\hline
\end{tabular}
\end{table*}

\section{Astrometry}
Coordinates in the $X,Y$ system of each CCD must be transformed onto a common system, such that objects 
can be identified and matched into a master list for subsequent proper-motion determinations.
For this first step, we use the S07 catalog, which gives equatorial coordinates derived using the 2MASS catalog
as astrometric reference. 
Each Subaru exposure is transformed into the S07 catalog using up to third order polynomials. 
Between a few hundred to a few thousand objects are used in these solutions, and the standard error of the 
transformations ranges between 50 to 70 mas in each coordinate.
We then combine
all equatorial position measurements at a given epoch to form an average catalog, keeping only those objects 
with at least 5 measurements.
Within each of the three epochs, we perform a new transformation of the Subaru exposure into the epoch's average catalog. 
In this transformation we use up to fourth-order polynomial field terms.
Note that the transformation into the average catalog is superior to that into the S07 catalog as
it 1) eliminates dispersion introduced by proper motions and the epoch differences with S07, and 
2) avoids limitations of the S07 astrometry, thus, representing best the precision of the Subaru data.
Also, by averaging the positions of an object as it falls on different chips and/or various regions of the same
chip, it is expected that position-dependent systematics will be somewhat alleviated. 
The effectiveness of this averaging in reducing such systematics will be limited by the number and
dither/offset pattern of the available archive observations. 
The standard errors of these intra-Subaru transformations range between 12 and 22 mas
for the 2005 and 2008 data, and up to $\sim 30$ mas for the 2001 data; whereas the
well-measured objects ($i = 19 - 22$) can reach a positional precision of $5-6$ mas. 
It is during this process that we realized the 2001 data
from CCD w9c2 were very poor, with an rms of $\sim 80$ mas in the DEC solution, and showing a distinct residual pattern 
along the $X$ coordinate of the CCD. For this reason we have discarded CCD-w9c2 data from further analysis.

\begin{figure*}
\includegraphics[scale=0.7,angle=-90]{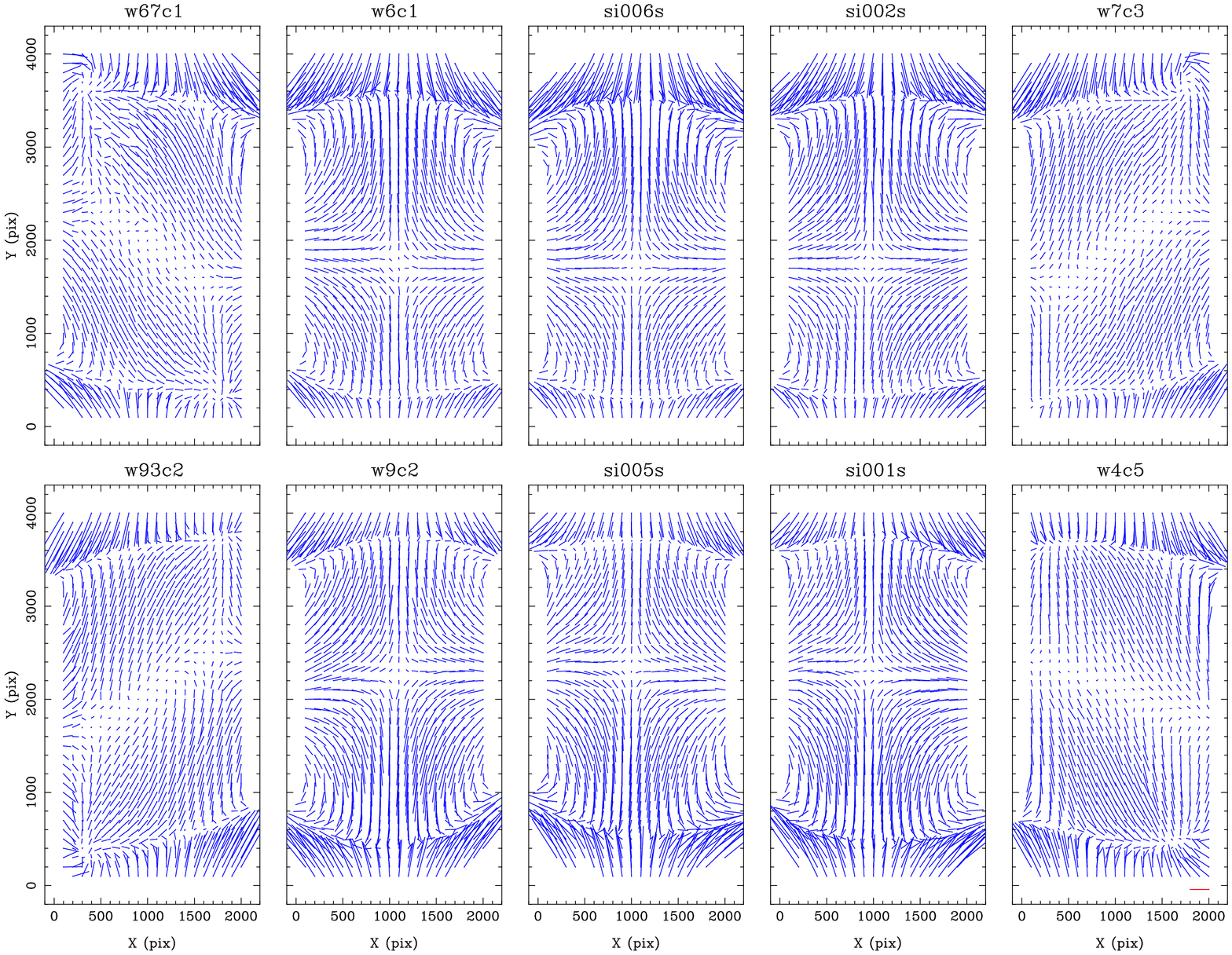}
\caption{Position residuals for the 2005 data exposures. Chip labels are as by Subaru convention. 
Residuals are stacked after 2nd-order polynomial coordinate transformations of each chip into an average 
catalog at this epoch (see text). Only well-measured objects ($i \sim 19-22$) are included. The scale of the residuals
is indicated by the red vector in the lower right corner of the last panel; this vector represents
a 100-mas residual.
}
\label{fig:fig5}
\end{figure*}   
\begin{figure*}
\includegraphics[scale=0.7,angle=-90]{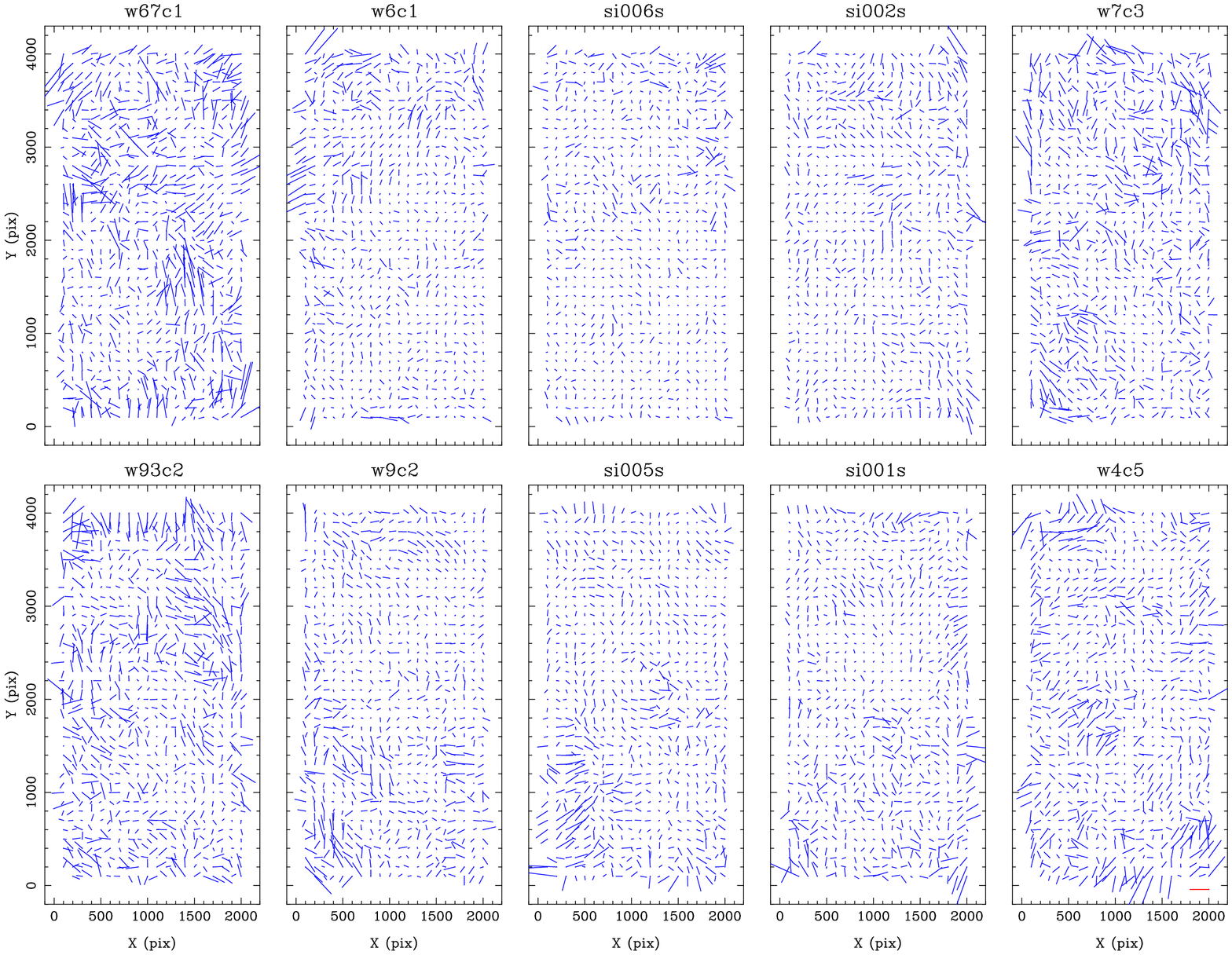}
\caption{Same as in Figure~\ref{fig:fig5}, but stacked residuals are after 4th-order polynomial 
transformations into an average 
catalog at the 2005 epoch. The scale of the residuals
is indicated by the red vector in the lower right corner of the last panel, this time representing
a 10-mas residual.
}
\label{fig:fig6}
\end{figure*}

In Figures~\ref{fig:fig5} and \ref{fig:fig6} we show the residual maps for the 2005 data, for each chip. 
To obtain these maps, we stack residuals 
of well-measured objects ($i \sim 19-22$) within spatial cells, using a grid composed of 10x10 CCD pixels per cell. 
For illustrative purposes, Figure~\ref{fig:fig5} shows residuals after 
a 2nd-order polynomial transformation, where the distortion pattern across the field of view is clearly apparent.
Figure~\ref{fig:fig6} shows the residuals after a 4th-order transformation, where the high-amplitude, lower order polynomials reflecting 
the global distortion pattern have been removed.
However, residual position systematics on small scales, within a chip, are still present. 
Also, higher-order residuals are apparent in the chips at the corners of the field of view. 
These systematics could be well-calibrated and corrected for, had the offset/dither pattern been designed with this 
purpose in mind. 
Since here we are limited to the available archived data, taken for purposes other than astrometry, we must
devise a procedure for extracting, as best as possible, the remaining positional systematics (see Section 5.2).

At this point in our procedure, each frame has positions placed on a common equatorial system and partially corrected 
for spatial variations across the chip, to the extent that the average catalog can account for these. 
These equatorial positions are
transformed via a gnomonic projection into standard $(\xi,\eta)$ coordinates with the origin at Draco's center (S07);
it is these coordinates that are used to determine proper motions. 

\subsection{Global Solution: Relative Proper Motions}

We begin our proper-motion determinations with a preliminary, relative proper-motion solution over the entire field.
This step is a practical convenience, as it provides precise proper motions
that are free of gross systematic errors.
The accuracy of the final proper motions is determined by the subsequent local solution with respect to background
galaxies, which is described in Section 5.2.
Still, a detailed description of the preliminary, global solution is warranted.

A total of 622 frames over the three epochs is used to construct a global, relative proper-motion solution.
Objects from all frames are collated into a master list with unique identifier, using a matching tolerance of $1\arcsec$.
Proper motions are determined only for objects that have at least 4 position measurements separated by a minimum  
of 3 years.

A classical plate-overlap solution is used to derive the star and galaxy parameters (positions and
proper motions in two coordinates), as well as the ``plate'' constants of each chip within each exposure \citep{eich88}.
These two different sets of unknowns can be solved for, simultaneously, via a large set of equations
constrained by least-squares minimization of residuals of reference stars into a common system, using the
reference-star positions at all epochs at which they are measured.
In practice, we use an iterative procedure that separates the chip-constant solutions from the proper-motion
determinations.
The reference star positions are updated to the epoch of each chip/exposure based on the previous iteration's
proper-motion estimates.
From these, new chip constants are derived for all chips and then, after applying these, new mean-epoch 
positions and proper motions are derived.
These new mean-epoch positions and proper motions provide the reference positions for the next iteration's
chip transformations.
Chip transformations consist of 4th-order polynomial field terms, using both stars and
galaxies as reference objects.
Relative proper motions and mean-epoch positions are determined by unweighted least-squares fitting 
of $\xi$ and $\eta$ as linear functions of time, for each object.
Proper-motion uncertainties are derived from the scatter about these linear fits.
Note that without an external anchor, the scale of the positions is unconstrained and could drift.
To counter this, after each iteration 
the positions of the current object catalog are compared to those of the S07 catalog and
an overall adjustment is made to keep the scale constant.
Likewise, spatial variations in the mean proper motion of the reference stars can accumulate.
Therefore, a second external constraint is applied, forcing the relative proper motion of the reference stars to be
zero across the field of study, up to fifth-order spatially.
As there is expected to be a gradient in the mean motion of the reference-star sample, due to the spatial
variation of Draco members, the relative proper-motion system may still contain low-order field terms.
These will be dealt with later;
our relative proper-motion scheme is intended to derive optimal precision from these astrometric data.

\begin{figure}
\includegraphics[width=\columnwidth,angle=0]{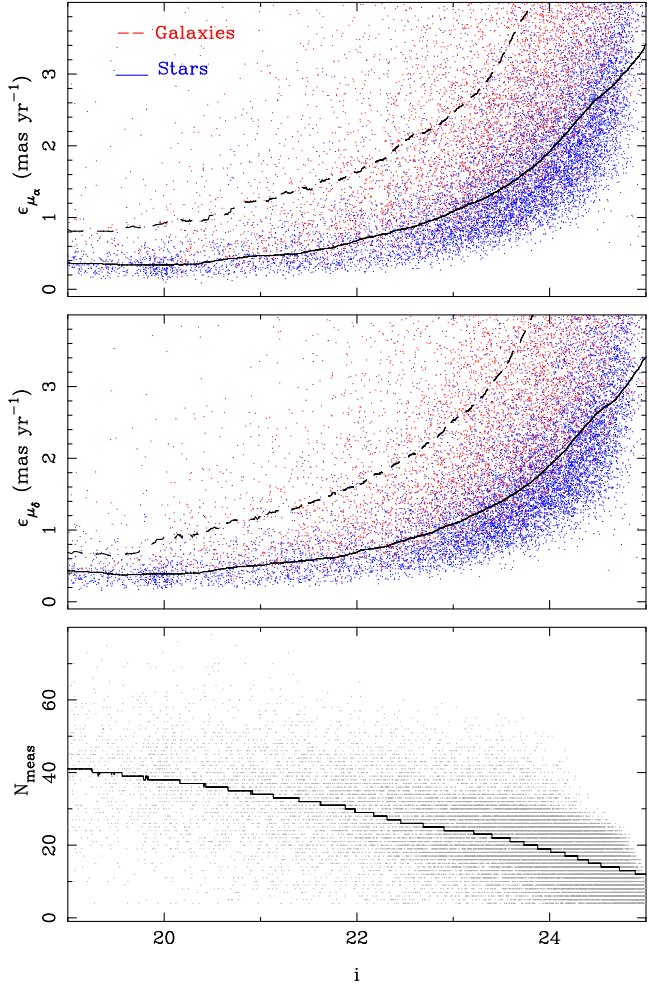}
\caption{Proper-motion uncertainties as a function of $i$ magnitude in each coordinate (top and middle panels)
for stars (blue) and galaxies (red). Moving medians for each are also shown as curves (solid for stars, and dashed for galaxies).
The bottom panel shows the 
number of positional measurements used in the proper-motion determinations as a function of $i$, (shown for $\mu_{\alpha}$, 
but it is similar for $\mu_{\delta}$). Once again, a moving median is shown as a solid curve.
}
\label{fig:fig7}
\end{figure} 

The relative proper-motion procedure converges after two or three iterations.
The resulting proper-motion uncertainties as a function of $i$ magnitude 
are shown in Figure~\ref{fig:fig7} for stars and galaxies.
Here, we also show the number of positions used in the proper-motion determination for each object 
(shown for $\mu_{\alpha}$, but it is similar for $\mu_{\delta}$)\footnote{throughout the paper $\mu_{\alpha}$ is actually $\mu_{\alpha}$~cos$\delta$, and $\mu_l$ is $\mu_l$~cos$b$}. 
Approximately $35\%$ of the objects in our proper-motion catalog are based on spans of 7 years (and three epochs), while
$65\%$ are from data spanning 4.4 years (two epochs). 
With roughly 30 to 40 position measurements for well-measured objects,
we obtain formal uncertainties of $\sim$ 0.4 to 0.5 mas~yr$^{-1}$ for stars with $19 < i < 21.4$. This corresponds
to a single-epoch positional error of 12 mas (using $\Delta t = 7$ years) for well measured stars, which is a
factor of $\sim 2$ larger than our position-error estimates in Section 4 (5-6 mas). The disagreement is likely due to the
presence of field-dependent systematics that vary from epoch to epoch, and which were not accounted for in our
modeling.
These systematics will be corrected in a subsequent ``local-solution'' adjustment described in the following section.
Galaxies have proper-motion uncertainties starting at about a factor of two larger than those of stars, with this
factor increasing toward fainter magnitudes.

\subsection{Local Solution: Absolute Proper Motions}

With relative proper motions in hand, determining the Draco dSph's 
motion with respect to an inertial reference frame is, in principle,
simple: it is the difference between the mean relative proper motion of Draco members
and that of distant background galaxies.
In practice, a number of complications arise and must be properly handled, as described below.
Nonetheless, the first step is to identify a set of Draco dSph member stars that can be used to determine
the system's mean motion.

\begin{figure}
\includegraphics[scale=0.7,angle=0]{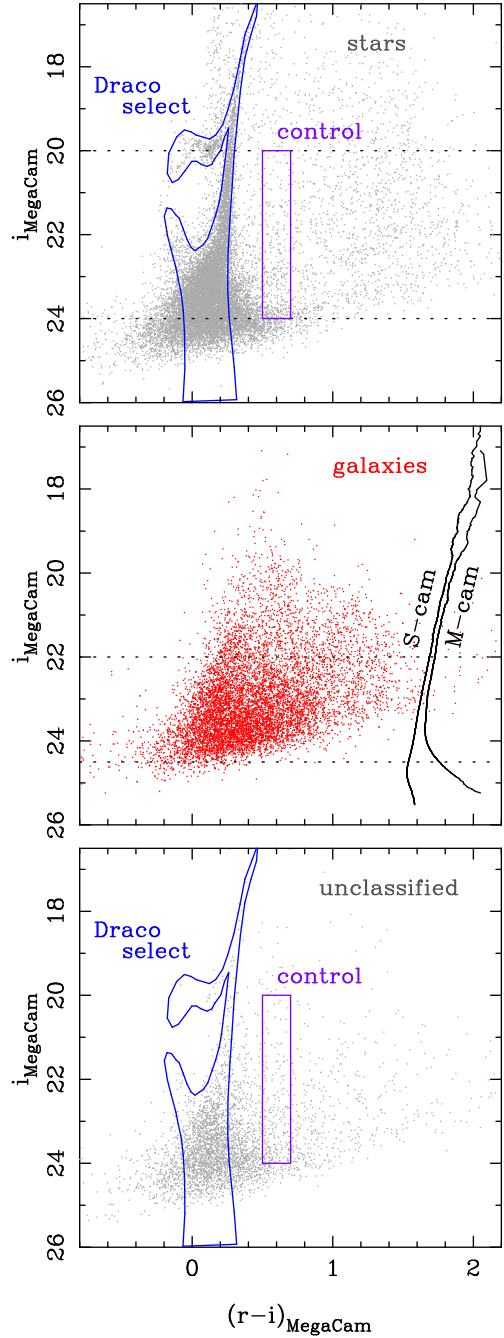}
\caption{Colour-magnitude diagrams of stars (top), galaxies (middle) and unclassified objects (bottom) for which proper motions are measured. The photometry is from S07. Stars and galaxies are from our classification (see Section 4). The Draco-star boundary is outlined, as is the rectangular box of our control field
used to estimate the field contamination in our Draco sample (see Section 5.3.3).
Dashed lines indicate the magnitude-range constraints applied to Draco stars and galaxies to be used in the analysis. 
The SuprimeCam observations we use are deeper than those of S07.  This is shown by the
$i$-magnitude marginal distributions of galaxies in our data and those cross-matched with S07
photometry, the two labeled curves on the right side of the middle panel. The scale is logarithmic and 
arbitrary, intended only to show that the SuprimeCam data is deeper. 
}
\label{fig:fig8}
\end{figure}

Draco members are selected photometrically from the $(r-i,i)$ colour-magnitude diagram (CMD).  To perform this selection,
we use only the photometry from S07.
In Figure~\ref{fig:fig8} we show the CMD using S07 photometry for stars, galaxies and unclassified objects, in separate panels.
A boundary for selecting the Draco sequence is also shown. In our proper-motion determination, we will
use only stars with $20 \le i \le 24$. Brighter than $i=20$, the astrometry is severely affected, while 
fainter than $i=24$, the confusion with the field stars may affect the mean proper motion of Draco. 
(The effect of contamination is estimated in Section 5.3.3.)  
The use of the S07 photometry solely for the selection
of Draco and control samples is advisable due to 
its better quality; however one should keep in mind that our data set 
is deeper, as indicated by the two magnitude distributions of galaxies in the middle
panel of Figure~\ref{fig:fig8}.  
For this reason, the object classification, based on the deeper SuprimeCam data, is found reliable 
down to $i=24.5$ (see Sections 4 and 5.3.1).


In the global, relative-proper-motion solution described above, reference objects for the inter-epoch
transformations were selected from all types of objects; Galactic foreground field stars, Draco stars, and 
background galaxies. 
As there is a variation in the ratio of these populations
across the field (especially in that of the Draco stars), 
the proper-motion reference system of our global solution also should be expected to vary. 
Thus, using an overall mean across the entire field-of-study is inappropriate. 
In addition, position-dependent systematics are likely to remain due to our inability
to perfectly model each chip at all epochs. 
To circumvent these issues, which might disturb the proper motions over a range of characteristic scale lengths, 
we perform a local solution as follows. 
For each Draco star, we choose the closest $N_n$ galaxies and determine the
median proper motion of these galaxies. 
At times, the target star might not be well-centered within these surrounding galaxies (e.g., near the edge
of the overall field), so an additional correction to this median value is made based on the slopes of the galaxies'
proper motions as a function of $\xi$ and $\eta$ and the offset between the target star's position and the mean of
the $N_n$ galaxies.
The proper motion of the Draco star is then corrected to absolute by subtracting the galaxies' adjusted median 
proper motion, which is a measure of the inertial frame at that position.
After several trials, the number of galaxies used to define the local inertial system was set to 100.
This choice yields a random uncertainty in the correction that is smaller than the estimated uncertainty in
relative proper motion, for a typical Draco star.
That is, $\epsilon_{\mu}^{gal}/\sqrt(100) \le \epsilon_{\mu}^{star}$, where 
$\epsilon_{\mu}^{gal} \sim 3-4 $ mas~yr$^{-1}$, and $\epsilon_{\mu}^{star} \sim 0.5 $ mas~yr$^{-1}$.
At the same time, choosing $N_n=100$ ensures that the size of the local inertial system is small compared to 
the size over which spatial variations are significant (see Section 5.3.1). 

\begin{figure*}
\includegraphics[scale=0.8,angle=-90]{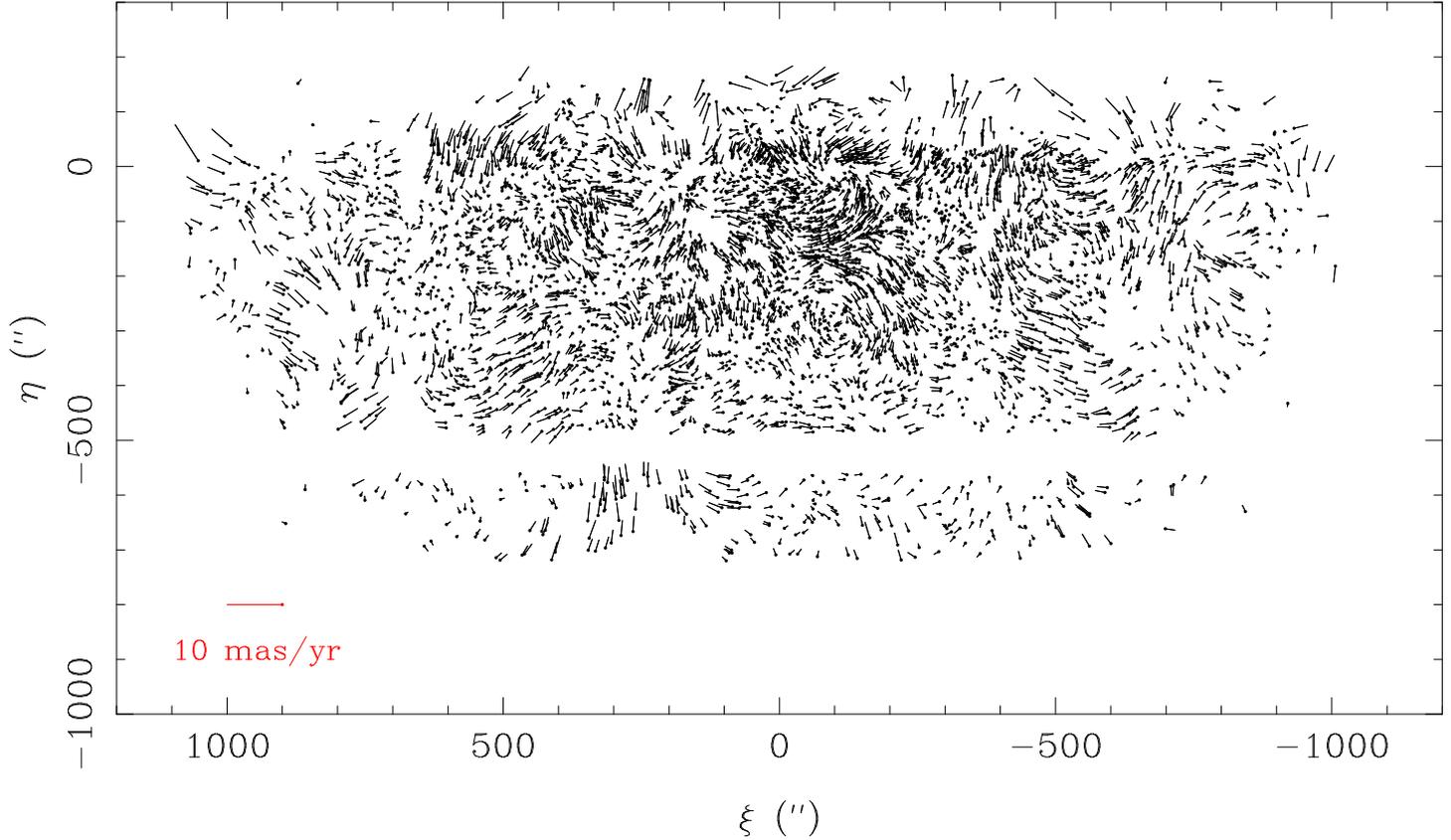}
\caption{Map of the median proper motion of the 100 galaxies closest to each Draco star, plotted
at the locations of the Draco stars that were used in the determination of the mean motion.
These are used as a local correction to absolute proper-motions (see text). The gap at
$\eta \sim -500''$ is due to the S07 CMD Draco member selection; the S07 data have a gap 
due to the arrangement of the chips on the MegaCam on CFHT. 
A vector of length 10 mas~yr$^{-1}$ is shown in red.}
\label{fig:fig9}
\end{figure*}

In Figure~\ref{fig:fig9} we show these local proper-motion corrections (i.e., the adjusted median proper motion of 
the 100 neighboring galaxies)
at the location of each Draco star used in the final determination of the system's motion. 
The systemic motion of Draco is determined from the unweighted mean of the CMD-selected Draco members.
We further restrict this set to stars that have 20 or more observations, and with proper-motion amplitude
$\sqrt(\mu_{\alpha}^2+\mu_{\delta}^2)   \le 5.0$ mas yr$^{-1}$ in order to remove non-Draco outliers.



\subsection{Exploring Possible Systematic Errors}

\subsubsection{Magnitude and Colour Trends for Galaxies} 
It is possible for us
to explore magnitude and color trends for the galaxies, given the unprecedented large number of galaxies used 
in this study ($\sim 12000$). In these tests, we fix the magnitude range for Draco stars to $i=20-24$ (see Section 5.3.2), while we vary
the reference galaxy sample. 
As a first test, we repeat the local-solution step using only galaxies from specific magnitude bins. 
Since this approach lowers the total number 
of galaxies participating in the solution, it is necessary to use a smaller number of galaxies per Draco star.
We test two values for $N_n$; 30 and 50 neighboring galaxies. These choices ensure that the mean radius of the 
local reference system remains comparable to that obtained in the nominal solution, i.e., $N_n$=100 for the full set of galaxies. 
In Table~\ref{tab:tab3} we list the absolute proper motion of Draco resulting from using each of these magnitude restricted
samples of galaxies for the local solution. 
The table lists the $i$ magnitude range; the number of Draco stars used in the solution, $N_D$; the 
total number of galaxies used in the given magnitude range, $N_g$; the average radius of the local reference system, $r$, 
in arcsec; and the choice for number of neighboring galaxies, $N_n$.
For each determination, the uncertainty in mean motion is from the standard deviation of the absolute proper motions of 
Draco stars divided by the square root of their number.
These are pure statistical errors which are underestimates, and are {\it not} used in our final estimation of the proper-motion uncertainty of Draco (see Section 5.4).

As seen in Table~\ref{tab:tab3}, the first and last magnitude bins yield
significantly different values from the other four magnitude bins and for this reason they are
excluded from the analysis.
It is possible to speculate as to why these limiting bins are unreliable.
At the bright end, both magnitude-dependent systematics, and the
small number of galaxies available for the local solution could contribute to a biased result; in this
case, towards large negative values. 
At the faint end, the values approach zero, which is indicative of contamination by field stars.
The mean absolute motion of the field stars is negative in both coordinates (see Section 5.3.3); mistakenly allowing field stars 
into the galaxy sample, produces an erroneously more positive mean motion for Draco. 
For these reasons, we will limit the galaxy sample to the range of the central magnitude bins, $i=22-24.5$.
The last lines in Table~\ref{tab:tab3} show the means of the four magnitude bins, for the solutions with 50 and 30 neighbors. 
The uncertainty in the mean is derived from the standard deviation of the four independent magnitude-bin 
determinations. 
Thus, this uncertainty is a better indicator of the residual systematic errors.
\begin{table*}
\caption{Absolute Proper-motion Solutions using Various Magnitude Bins for Galaxies}
\label{tab:tab3}
\begin{tabular}{lllllll}
\hline
\multicolumn{1}{c}{$i$ range} & \multicolumn{1}{c}{$\mu_{\alpha}$} & \multicolumn{1}{c}{$\mu_{\delta}$} & \multicolumn{1}{c}{$N_D$} & \multicolumn{1}{c}{$N_g$} & \multicolumn{1}{c}{$r$} & \multicolumn{1}{c}{$N_n$} \\
 & \multicolumn{1}{c}{(mas yr$^{-1}$)} & \multicolumn{1}{c}{(mas yr$^{-1}$)} & & & ($\arcsec$) & \\
\hline
\hline
\multicolumn{1}{l}{20-22} & \multicolumn{1}{r}{\it $-0.701\pm0.029^{\dagger}$} & \multicolumn{1}{r}{\it $-0.514\pm0.028$} & 4253  & 1664 & 97 & 50 \\
\multicolumn{1}{l}{20-22} & \multicolumn{1}{r}{\it $-0.674\pm0.031$} & \multicolumn{1}{r}{\it $-0.548\pm0.029$} & 4153  & 1664 & 75 & 30 \\ 
\multicolumn{7}{c}{.................................................................................................................} \\
\multicolumn{1}{l}{22-23} & \multicolumn{1}{r}{$-0.186\pm0.028$} & \multicolumn{1}{r}{$-0.173\pm0.028$} & 4333  & 2425 & 75 & 50 \\ 
\multicolumn{1}{l}{22-23} & \multicolumn{1}{r}{$-0.196\pm0.029$} & \multicolumn{1}{r}{$-0.201\pm0.030$} & 4271  & 2425 & 59 & 30 \\ \\
\multicolumn{1}{l}{23-23.5} & \multicolumn{1}{r}{$-0.288\pm0.028$} & \multicolumn{1}{r}{$-0.236\pm0.029$} & 4318  & 2304 & 77 & 50 \\ 
\multicolumn{1}{l}{23-23.5} & \multicolumn{1}{r}{$-0.278\pm0.030$} & \multicolumn{1}{r}{$-0.257\pm0.030$} & 4247  & 2304 & 60 & 30 \\ \\
\multicolumn{1}{l}{23.5-24} & \multicolumn{1}{r}{$-0.135\pm0.030$} & \multicolumn{1}{r}{$-0.220\pm0.028$} & 4261  & 3054 & 68 & 50 \\ 
\multicolumn{1}{l}{23.5-24} & \multicolumn{1}{r}{$-0.140\pm0.031$} & \multicolumn{1}{r}{$-0.226\pm0.028$} & 4261  & 3054 & 53 & 30 \\ \\
\multicolumn{1}{l}{24-24.5} & \multicolumn{1}{r}{$-0.320\pm0.031$} & \multicolumn{1}{r}{$-0.306\pm0.031$} & 4162  & 3873 & 62 & 50 \\ 
\multicolumn{1}{l}{24-24.5} & \multicolumn{1}{r}{$-0.312\pm0.033$} & \multicolumn{1}{r}{$-0.305\pm0.033$} & 3953  & 3873 & 49 & 30 \\ 
\multicolumn{7}{c}{.................................................................................................................} \\
\multicolumn{1}{l}{24.5-25} & \multicolumn{1}{r}{\it $-0.002\pm0.033$} & \multicolumn{1}{r}{\it $-0.046\pm0.033$} & 3965  & 3157 & 70 & 50 \\ 
\multicolumn{1}{l}{24.5-25} & \multicolumn{1}{r}{\it $-0.013\pm0.037$} & \multicolumn{1}{r}{\it $-0.100\pm0.036$} & 3965  & 3157 & 55 & 30 \\ 
\hline 
\hline
\multicolumn{1}{l}{ave50} & \multicolumn{1}{r}{$-0.232\pm0.043^{\ddagger}$} & \multicolumn{1}{r}{$-0.234\pm0.028$} & & & &  \\
\multicolumn{1}{l}{ave30} & \multicolumn{1}{r}{$-0.231\pm0.039$} & \multicolumn{1}{r}{$-0.247\pm0.022$} & & & &  \\
\hline
\multicolumn{7}{l}{{$\dagger$} Uncertainties for each magnitude bin are only the statistical component } \\
\multicolumn{7}{l}{due to scatter in the Draco stars; other sources of error are present.} \\
\multicolumn{7}{l}{{$\ddagger$} Uncertainties in the averages are derived from the
bin-to-bin scatter.} \\
\end{tabular}
\end{table*}

In a similar manner, we explore local solutions using two samples of galaxies binned by $(r-i)$ colour. 
(Uncertainties in the $(r-i)$ colour values precludes making a finer division by colour.)
These solutions are 
presented in Table~\ref{tab:tab4}, the columns being similar to those given in Table~\ref{tab:tab3}.
As the colour-binned samples are larger than in the tests by magnitude bin, we perform only solutions using the 50
closest galaxy neighbors. 
The two colour bins give mean motions for Draco that agree with one another.
Reassuringly, the average of the two bins' results agrees within the uncertainties with
that of the magnitude-bin tests. 
\begin{table*}
\caption{Absolute Proper-motion Solutions using Two Colour Bins for Galaxies}
\label{tab:tab4}
\begin{tabular}{lllllll}
\hline
\multicolumn{1}{c}{$(r-i)$} & \multicolumn{1}{c}{$\mu_{\alpha}$} & \multicolumn{1}{c}{$\mu_{\delta}$} & \multicolumn{1}{c}{$N_D$} & \multicolumn{1}{c}{$N_g$} & \multicolumn{1}{c}{$r$} & \multicolumn{1}{c}{$N_n$} \\
\multicolumn{1}{c}{range} & \multicolumn{1}{c}{(mas yr$^{-1}$)} & \multicolumn{1}{c}{(mas yr$^{-1}$)} & & & ($\arcsec$) & \\
\hline
\hline
\multicolumn{1}{l}{$\le 0.4$} & \multicolumn{1}{r}{ $-0.306\pm0.029^{\dagger}$} & \multicolumn{1}{r}{$-0.258\pm0.027$} & 4358  & 6264 & 46 & 50 \\
\multicolumn{1}{l}{$\ge 0.4$} & \multicolumn{1}{r}{ $-0.255\pm0.029$} & \multicolumn{1}{r}{$-0.201\pm0.029$} & 4298  & 5544 & 55 & 50 \\
\hline 
\hline
\multicolumn{1}{l}{ave} & \multicolumn{1}{r}{$-0.281\pm0.026^{\ddagger}$} & \multicolumn{1}{r}{$-0.230\pm0.029$} & & & &  \\
\hline
\multicolumn{7}{l}{{$\dagger$} Uncertainties in the two colour bins are only the component due to} \\
\multicolumn{7}{l}{scatter in each bin's Draco star motions.} \\
\multicolumn{7}{l}{{$\ddagger$} Uncertainties in the average is from the bin differences.} \\
\end{tabular}
\end{table*}

We also test for sensitivity to the choice of $N_n$, the number of reference galaxies used in the local solution.
For this test, we use the entire set of galaxies, with $i=22-24.5$. Seven values of $N_n$ are explored and the results
listed in Table~\ref{tab:tab5}.
The standard deviation of these solutions is $\sigma_{\mu_{\alpha}} = 0.020$ mas yr$^{-1}$ and 
$\sigma_{\mu_{\delta}} =  0.026$  mas yr$^{-1}$, making them consistent with the formal errors of each solution. 
The solutions are also in reasonable agreement with values obtained from the magnitude and colour tests.
\begin{table}
\caption{Absolute Proper-motion Solutions using Various Sizes of the Local Reference System}
\label{tab:tab5}
\begin{tabular}{rrrrrr}
\hline
 \multicolumn{1}{c}{$\mu_{\alpha}$} & \multicolumn{1}{c}{$\mu_{\delta}$} & \multicolumn{1}{c}{$N_D$} & \multicolumn{1}{c}{$N_g$} & \multicolumn{1}{c}{$r$} & \multicolumn{1}{c}{$N_n$} \\
\multicolumn{1}{c}{(mas yr$^{-1}$)} & \multicolumn{1}{c}{(mas yr$^{-1}$)} & & & ($\arcsec$) & \\
\hline
\multicolumn{1}{r}{ $-0.253\pm0.039^{\dagger}$} & \multicolumn{1}{r}{$-0.244\pm0.037$} & 3297  & 12329 & 15 & 10 \\
\multicolumn{1}{r}{ $-0.259\pm0.031$} & \multicolumn{1}{r}{$-0.280\pm0.030$} & 4184  & 12329 & 27 & 30 \\
\multicolumn{1}{r}{ $-0.310\pm0.031$} & \multicolumn{1}{r}{$-0.296\pm0.030$} & 4329  & 12329 & 34 & 50 \\
\multicolumn{1}{r}{ $-0.284\pm0.028$} & \multicolumn{1}{r}{$-0.289\pm0.028$} & 4413  & 12329 & 49 & 100 \\
\multicolumn{1}{r}{ $-0.269\pm0.027$} & \multicolumn{1}{r}{$-0.258\pm0.028$} & 4421  & 12329 & 69 & 200 \\
\multicolumn{1}{r}{ $-0.255\pm0.027$} & \multicolumn{1}{r}{$-0.240\pm0.028$} & 4423  & 12329 & 84 & 300 \\
\multicolumn{1}{r}{ $-0.264\pm0.027$} & \multicolumn{1}{r}{$-0.231\pm0.028$} & 4421  & 12329 & 97 & 400 \\
\hline
\multicolumn{6}{l}{{$\dagger$} Uncertainties within each bin are only the component due to} \\
\multicolumn{6}{l}{scatter in the Draco star motions.} \\
\end{tabular}
\end{table}

Blended binary stars can be mistaken for galaxies and thus potentially bias our galaxy sample. 
Field binaries will have the same mean 
absolute motion as field single stars.
This is shown in Sect 5.3.3 to be
negative in both coordinates, from a comparison to the Besancon model \citep{rob03}, as well as from our data.  
Contamination of the galaxy sample by field binaries would produce a bias toward a more positive proper motion of Draco, 
the same effect as it would be with single-star contamination.
Nevertheless, we now attempt to estimate the amount of contamination and bias by field binaries in
the magnitude range $i=20 - 22$.  We use the binary statistics of \citet{ragh10} considering
spectral types F6 to K4, which approximately corresponds to the stellar population under discussion.  We assume a
binary fraction of 0.3, with mass ratios between 0.7 and 1.0 ($\Delta~V=2.5$ to 0).  Our median
seeing of $0\farcs54$ and a/b ratio limit of 2 for reference galaxies implies an upper limit of $1\farcs08$ for
the separation of a binary that could masquerade as a galaxy.  As a lower limit for the separation,
we assume the binary becomes distinguishable from a single star at a separation of half the seeing,
or $0\farcs27$.  Within the magnitude range $i=20-22$, the Besancon model predicts that the average distance
to field stars is 5 kpc.  Translating these separation limits from arcsec to AU at this distance, and 
using the known separation distribution of binaries from Fig. 13 in \citet{ragh10} we estimate
that $8.4\%$ of the field binaries fall within these limits, or $2.5\%$ of the total number of field stars.
Scaling Besancon counts to our data, we obtain an estimate of 26 binaries (within $i=20-22$) that
could potentially contaminate our 1664 galaxies in this magnitude range, or $1.6\%$ contamination.
Note that this is an over-estimate to the extent that no proper-motion trimming has been applied to
remove large proper-motion outliers, as was done in our actual galaxy and star samples.  As the mean
motion of the field is roughly -0.5 mas/yr in both coordinates (see Sect 5.3.3), the estimated bias
is of the order 0.008 mas/yr, well below other errors.
A similar estimate for the second brightest bin, from $i=22-23$, yields a value of 
contamination of galaxies by field binaries of just $0.5\%$.

On the other hand, if Draco binaries are contaminating the galaxy sample in the range $i=20-22$, the
effect would be to pull the derived Draco absolute motion closer to zero.  This is not at all what
is seen for this magnitude bin.  Although, as is pointed out for single-star contamination of the
faintest magnitude bin (which is rejected), Draco binary-star contamination is consistent with the
deviation seen in the $i=24.5-25$ bin.

Finally, we consider a set of 11 quasars from the Half Million Quasar (HMQ) catalogue \citep{fle15} 
that are present in our proper-motion catalog. These are at magnitudes brighter than $i=21$, and thus prone to magnitude bias.
In any event, their mean motion with respect to galaxies has an uncertainty of $\sim 0.5$ mas yr$^{-1}$, 
rendering them inadequate to serve as possible probes of systematics.

\subsubsection{Magnitude and Color Trends for Draco Stars}

\begin{figure}
\includegraphics[width=\columnwidth,angle=0]{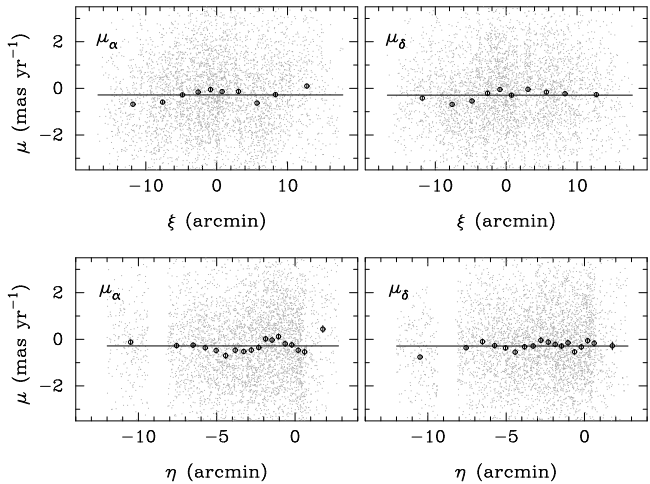}
\caption{Absolute proper motions of Draco stars as a function of $\xi$ and $\eta$. Large symbols
represent averages in bins of equal number of stars. Error bars (included in the plot) are about the size of these
symbols. The horizontal lines show the averages for the
entire sample. 
}
\label{fig:fig10}
\end{figure}

Having established the valid magnitude range for galaxies from $i=22-24.5$, we now test 
solutions based on various magnitude bins for the Draco sample. We use a local reference system of 100 galaxies and
a total number of galaxies of 12329, resulting in an average radius of the reference system of 49$\arcsec$ (see Section 5.3.1).
Results are listed in Table~\ref{tab:tab6}. 
The brightest magnitude bin produces an obvious outlier mean motion and we discard it.
The faintest magnitude bin is also excluded, not because the result indicates a magnitude bias, 
but because
the photometry used to isolate Draco members via the CMD becomes less reliable at the faint end of the S07 study, 
producing a large statistical error compared to the other magnitude bins. 
Results from the remaining six bins show scatter, but they do not indicate a trend with magnitude. 
The average of these six magnitude bins is presented in the last line of Table~\ref{tab:tab6}, with an uncertainty based on
the standard deviation of the six values.
We have also plotted the Draco stars' proper motions as a function of $(r-i)$ colour, and find no significant trends.
\begin{table}
\caption{Absolute Proper-motion Solutions using Various Magnitude Bins for Draco Stars}
\label{tab:tab6}
\begin{tabular}{rrrr}
\hline
 \multicolumn{1}{c}{$i$ range} &\multicolumn{1}{c}{$\mu_{\alpha}$} & \multicolumn{1}{c}{$\mu_{\delta}$} & \multicolumn{1}{c}{$N_D$} \\
 & \multicolumn{1}{c}{(mas yr$^{-1}$)} & \multicolumn{1}{c}{(mas yr$^{-1}$)} &  \\
\hline
\hline
\multicolumn{1}{l}{19-20} & \multicolumn{1}{r}{ $-0.002\pm0.068^{\dagger}$} & \multicolumn{1}{r}{$-0.071\pm0.063$} & 419   \\
\multicolumn{4}{c}{...............................................................................................} \\
\multicolumn{1}{l}{20-21} & \multicolumn{1}{r}{ $-0.118\pm0.066$} & \multicolumn{1}{r}{$-0.247\pm0.063$} & 440  \\
\multicolumn{1}{l}{21-22} & \multicolumn{1}{r}{ $-0.371\pm0.072$} & \multicolumn{1}{r}{$-0.186\pm0.070$} & 511   \\
\multicolumn{1}{l}{22-22.5} & \multicolumn{1}{r}{ $-0.195\pm0.089$} & \multicolumn{1}{r}{$-0.231\pm0.091$} & 417  \\
\multicolumn{1}{l}{22.5-23} & \multicolumn{1}{r}{ $-0.277\pm0.066$} & \multicolumn{1}{r}{$-0.259\pm0.061$} & 749   \\
\multicolumn{1}{l}{23-23.5} & \multicolumn{1}{r}{ $-0.285\pm0.055$} & \multicolumn{1}{r}{$-0.309\pm0.052$} & 1158   \\
\multicolumn{1}{l}{23.5-24} & \multicolumn{1}{r}{ $-0.341\pm0.055$} & \multicolumn{1}{r}{$-0.378\pm0.052$} & 1143   \\
\multicolumn{4}{c}{...............................................................................................} \\
\multicolumn{1}{l}{24.0-24.5} & \multicolumn{1}{r}{ $-0.306\pm0.081$} & \multicolumn{1}{r}{$-0.378\pm0.081$} & 672   \\
\hline
\hline
\multicolumn{1}{l}{ave20-24 bins} & \multicolumn{1}{r}{ $-0.265\pm0.038^{\ddagger}$} & \multicolumn{1}{r}{$-0.268\pm0.027$} & \\
\hline
\multicolumn{4}{l}{{$\dagger$} Uncertainties listed for each magnitude bin are only the} \\
\multicolumn{4}{l}{component due to scatter in the bin's Draco star motions.} \\
\multicolumn{4}{l}{{$\ddagger$} Uncertainties in the averages are from bin-to-bin scatter.} \\
\end{tabular}
\end{table}
Finally, we show the absolute proper motions of Draco stars as a function of spatial coordinates $(\xi, \eta)$ in Figure~\ref{fig:fig10}.
This is the solution with 100 neighboring galaxies in the local 
reference system, using only galaxies within $22 \le i \le 24.5$ and Draco stars within $20 \le i \le 24$, i.e., the 
fourth-line solution in Table~\ref{tab:tab5}. 
Averages in bins of equal number of stars are also shown (large symbols), together with the average for the entire sample (continuous
line). No global significant trend is seen in these plots.

Inspecting Tables~\ref{tab:tab3} through \ref{tab:tab6}, we find that it is the
solutions as a function of magnitude, for both galaxies and Draco stars, that show the largest scatter, in comparison to
the other tests. Also, the $\mu_{\alpha}$ scatter is larger than the $\mu_{\delta}$ scatter. 
Still, these values of scatter correspond to 
errors of the order of $\sim 0.04$ mas yr$^{-1}$ in $\mu_{\alpha}$, and $\sim 0.03$ mas yr$^{-1}$ in $\mu_{\delta}$
when the ensemble of useful magnitude ranges is considered.

\subsubsection{Stellar Field Contamination of the Draco Sample}
Now, we estimate the proper-motion bias introduced by the contamination of the Draco sample by field stars.
The Besancon model \citep{rob03} is used to simulate the properties of the Galactic stars at the location of our field, 
over an area ten times larger, for better statistics. 
Thus, we assume the mean motions of various stellar populations do not change considerably
over an area of $\sim 1\fdg5$. The $(r-i,i)$ CMD is used to isolate two samples for comparison, in both the 
observations and in the model. 
The first sample is defined by a boundary for the Draco stellar sequence
used in Section 5.2.1, and within $20 \le i \le 24$. The second sample is a control field, also
in the magnitude range $20 \le i \le 24$ but offset to the 
color range $0.5 \le (r-i) \le 0.7$. By comparing star counts in the ``on-Draco'' and ``control'' samples in the model and then
scaling to the counts in the observations, we can estimate the foreground contamination fraction in our Draco sample.  
As the Draco sample in our study was trimmed in proper motion
to exclude objects with proper motions $> 0.5$ mas yr$^{-1}$, we apply a similar cut to all samples considered in this exercise.
This proper-motion cut has to be done carefully however, as the mean
motion of the control field in the model data is not close to zero as in our observations.
First, we must align the two proper-motion
systems (observations and model) using an offset determined from the control-field samples. 
Afterward, all samples can then be given the proper-motion cut.
The ratio of stars in the on-Draco and control samples of the model are used to predict the number of field stars
in the Draco observations sample, which is then expressed as a fractional contamination.
The results is an estimated $3.5\pm0.2\%$ contamination, where the uncertainty is from Poisson statistics of the real data.
This fraction might seem small considering possible confusion at the faint end of the samples, but keep in mind that above
the faint limit of $i=24$, the turnoff and main sequence 
of Draco contribute large numbers of stars that will dominate over the Galaxy's foreground halo.

The field-contamination fraction times the mean motion of the field stars gives an estimate of the bias introduced by
field stars in our actual Draco sample.
We could simply use the Besancon proper motions in the on-Draco sample to predict this bias. However, 
we prefer to work in a differential way, rather than relying on the absolute kinematics from the model predictions.
We calculate the mean motions of stars in both the on-Draco and control samples and determine the offset 
between these two samples. This offset is then applied to the mean motion of the control sample of the observations, to
predict the mean motion of field stars in the Draco sample. Finally, this predicted motion is scaled by the contamination 
fraction.
In this manner, we obtain a proper-motion bias $(\mu_{\alpha},\mu_{\delta}) = (-0.001\pm0.007, 0.003\pm0.007)$ mas yr$^{-1}$. 
This bias should be subtracted from our Draco proper motion, i.e., $\mu_{\alpha}$ would become more positive, and
$\mu_{\delta}$ more negative.
 
Another way to constrain the bias, in terms of an upper limit, is as follows.
The mean motion of the field as predicted by the Besancon model in the control sample is more negative (larger in absolute value)
than the mean motion of the on-Draco sample; this is because, on the mean, stars in the on-Draco sample are more distant than
stars in the control field. If we take the mean motion obtained from the observations in the control field  $(\mu_{\alpha},\mu_{\delta}) = (-0.495\pm0.199, -0.559\pm0.187)$ mas yr$^{-1}$ and assume it to be the same as in the on-Draco field,
we obtain a proper-motion bias of $(\mu_{\alpha},\mu_{\delta}) = (-0.017\pm0.007, -0.020\pm0.007)$ mas yr$^{-1}$.
As just noted, the Besancon model shows that the control field is actually more negative, and therefore 
this value represents an upper limit.
In both cases our estimate of the bias is much less than our formal errors in the mean motion of Draco,
 and thus we decide not to apply it.

\subsection{Final Draco Proper Motion}
We adopt as the final Draco proper-motion value the solution with 100 neighboring galaxies in the local 
reference system, using only galaxies within $22 \le i \le 24.5$ and Draco stars within $20 \le i \le 24$, i.e., the 
fourth-line solution in Table~\ref{tab:tab5}. 
The total uncertainty in the mean absolute motion of Draco will have a component from the errors in the relative
proper motions of Draco members and a component from the uncertainty in the absolute reference frame, as determined
by the errors in the relative proper motions of the galaxies.
The uncertainties listed in Table~\ref{tab:tab5} were calculated from the scatter of the galaxy-corrected Draco star
proper motions divided by the number of Draco stars; that is, as if the scatter were due entirely to the first source
of uncertainty.
It is necessary, though, to also estimate the contribution from the second component, that associated with the galaxies.
In fact, the two effects can be deduced from the variation in observed scatter of corrected Draco star motions over the
various trial solutions in Table~\ref{tab:tab5}.
We use the trend of these formal errors with the number of galaxies used in the local solution, $N_g$,
to determine the average proper-motion error for stars and for galaxies, from which the overall uncertainty in
the mean absolute motion of Draco can be derived.

While proper-motion errors for individual stars and galaxies could be read from Figure~\ref{fig:fig7}, what is needed is the 
average error of each type over the magnitude range used.
These are better estimated empirically, from the scatter of the actual galaxy-corrected proper motions of Draco stars.
Let $\sigma_s$ and $\sigma_g$ be the average, relative proper-motion errors for a Draco star and
for a galaxy, respectively. 
Then, the individual error in the proper motion of a Draco star after applying the 
local correction is $\sigma_o^{2} = \sigma_s^2 + \sigma_g^2/N_g $. 
The quantity listed in Table~\ref{tab:tab5} as an uncertainty is, in fact,
$\epsilon = \sigma_o/\sqrt(N_D)$, where $N_D$ is the number of Draco stars used in the solution. 
Thus, $\epsilon^2 \times N_D = \sigma_o^2 = \sigma_s^2 + \sigma_g^2/N_g$. Plotting $\epsilon^2 \times N_D$ as a function of
$1/N_g$ for the seven tests in Table~\ref{tab:tab5}, we fit a line to determine the intercept and the slope, i.e.,
the individual average proper-motion errors for stars and galaxies. We find $\sigma_s^{\mu_{\alpha}} = 1.82$ and 
$\sigma_s^{\mu_{\delta}} = 1.89$ mas yr$^{-1}$, and for galaxies $\sigma_g^{\mu_{\alpha}} = 4.27$ and 
$\sigma_g^{\mu_{\delta}} = 3.29$ mas yr$^{-1}$. 
We note that the observed scatter will also have been inflated by the presence of any 
magnitude-dependent or color-dependent variations, as seen in Tables~\ref{tab:tab3},\ref{tab:tab4} and
\ref{tab:tab6}, since the entire magnitude and color ranges were included in the solutions of Table~\ref{tab:tab5}.

While the derived errors appear large, this is not too surprising, considering that our
samples are dominated by faint objects. Using these values we can now calculate the uncertainty in the 100-neighboring galaxies 
solution. Our final proper motion estimate for Draco is, thus,
$(\mu_{\alpha}, \mu_{\delta}) = (-0.284\pm0.047, -0.289\pm0.041)$ mas~yr$^{-1}$. 
This estimate agrees, within the uncertainties,
with the average estimates over magnitude and color bins presented in Tables~\ref{tab:tab3},\ref{tab:tab4} and
\ref{tab:tab6}, where various systematic trends were explored. 
This provides evidence that magnitude and color-related systematics, to the extent they are present,
are at a level below our estimated uncertainty.

\subsection{Comparison with the $HST$ Determination}
In a recent study, P15 measured the absolute proper motion of the Draco dSph 
using $HST$ imaging and obtained $(\mu_{\alpha},\mu_{\delta}) = (0.177\pm0.063, -0.221\pm0.063)$ mas yr$^{-1}$. 
The P15 study used a repeated pointing with the Wide Field Channel (WFC) of the Advanced Camera for Surveys (ACS),
over a time baseline of 2 years. 
The WFC has two fields, WFC1 and WFC2.
Within each of these fields, one QSO was identified and used as a reference object. 
Also, compact background galaxies were used separately as reference in both fields (about 100 galaxies per field), the study thus providing four distinct 
measurements of the motion of Draco. 

\begin{figure}
\includegraphics[width=\columnwidth,angle=-90]{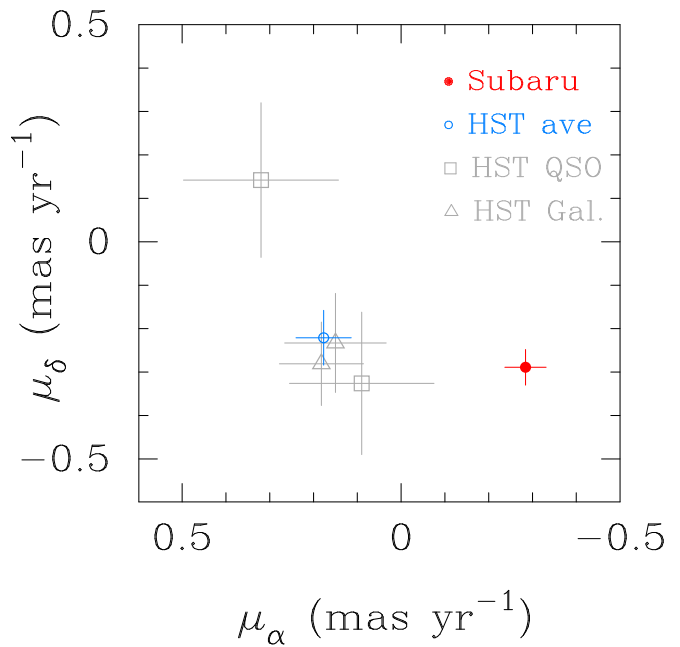}
\caption{Estimates of the absolute proper motion of the Draco dSph. 
$HST$ determinations are shown with
open symbols: triangles for averages with respect to background galaxies, squares for measures with
respect to individual QSOs, and a blue circle to indicate the 
weighted mean of the four determinations, as adopted by P15 to represent their overall $HST$ result.
Our measurement, based on Subaru observations, is shown with a red symbol.
Error bars are 1-$\sigma$ uncertainties.
}
\label{fig:fig11}
\end{figure}

In Figure~\ref{fig:fig11} we show these four determinations, together with the final 
result adopted by P15, a weighted mean of the 
four measures (open symbols). 
In the same plot we show our result (filled symbol). 
While in $\mu_{\delta}$ the determinations agree within estimated uncertainties, 
in $\mu_{\alpha}$ they are discrepant at the $\sim 6\sigma$ level. 

The source of the disagreement in $\mu_{\alpha}$ is unclear.
For the SuprimeCam detectors, the readout direction is along the $y$ direction, 
which in the Draco data set is always along $\delta$.
We have performed other tests, based on proper-motion solutions obtained from simple chip pairs
at two different epochs (2001-2005, and 2001-2008); we consistently obtain a negative value for $\mu_{\alpha}$
of Draco.
Another trial was made including two photographic plates taken with the 4m Mayall telescope in 1974.
These were measured with the Yale microdensitometer and reduced with the procedure described in \citet{cd06}.
The plate measures were combined with one Subaru exposure each from the 2001 and 2005 data, thus 
providing a $\sim 30$-year baseline to check the proper motion. 
The resulting Draco proper motion maintains the same 
orientation as obtained from the Subaru data alone. 
We do not describe this trial reduction in more detail as it is 
of lower accuracy.  However, we note that in all of our various solutions, the motion of Draco $\mu_{\alpha}$
is negative.

Since, the discrepancy in $\mu_{\alpha}$ is roughly a sign reversal, we have double-checked the
orientation in both our study and, to the extent possible, in the P15 study.
These appear correct and, thus, transformation errors are not the cause of the discrepancy.
We do note, however, that the largest scatter in the four proper-motion measurements of P15
(their Figure 13) is along the y-axis of the WFC/ACS detector, i.e., along the direction
in which CTE correction was applied to the HST data.

\section{Space Velocity and Orbit}

To calculate its space velocity, we adopt the following parameters for Draco:
$(\alpha,\delta)_{(J2000)}$ = ($17^h$:$20^m$:$18.1^s$, $57\degr$:$55\arcmin$:$13\arcsec$) \citep{pia02},
heliocentric distance $82.4\pm5.8$ kpc \citep{kin08}, and
heliocentric radial velocity $-293.3\pm1.0$ km~s$^{-1}$ 
\citep{ar95}. The peculiar solar motion is
$(U_{\odot},V_{\odot},W_{\odot}) = 
(-10.0\pm0.36,5.25\pm0.62,7.17\pm0.38)$ km~s$^{-1}$ 
\citep{db98},
with positive $U$ radially away from the Galactic center, $V$ in the direction of Galactic rotation, 
and $W$ toward the North Galactic Pole. 
The Local Standard of Rest velocity is 237 km~s$^{-1}$, and the Sun's distance
to the Galactic center is $R_{0} = 8.0$ kpc. 
All these parameters
were purposely chosen the same as in P15, in order to make a 
straightforward comparison with their velocities. 
While the true velocity of the LSR might be slightly lower 
(e.g.,  232 km~s$^{-1}$, \citet{car12,bo12}), this will not affect our conclusions here. 

The proper-motion estimates are first transformed to the Galactic rest frame 
by subtracting the solar reflex motion at the distance of Draco. 
Galactic rest-frame values are listed in Table~\ref{tab:tab7}, in
celestial and galactic coordinates. The first line of the Table shows the P15 result, the second line, ours.
\begin{table*}
\caption{Galactic rest-frame proper motions}
\label{tab:tab7}
\begin{tabular}{lrrrr}
\hline
\multicolumn{1}{c}{Measurement} & \multicolumn{1}{c}{$\mu_{\alpha}^{grf}$} &  \multicolumn{1}{c}{$\mu_{\delta}^{grf}$} & \multicolumn{1}{c}{$\mu_{l}^{grf}$} & \multicolumn{1}{c}{$\mu_{b}^{grf}$} \\
 & \multicolumn{2}{c}{(mas~yr$^{-1}$)} & \multicolumn{2}{c}{(mas~yr$^{-1}$)} \\
\hline
P15 & $0.514\pm0.063$ & $ -0.187\pm0.063$ & $-0.218\pm0.063$ & $-0.501\pm0.063$ \\
Subaru & $0.053\pm0.047$ & $ -0.254\pm0.041$ & $-0.257\pm0.041$ & $-0.037\pm0.047$ \\
\hline
\end{tabular}
\end{table*}

Table~\ref{tab:tab7} shows that our proper motion has a smaller magnitude than that of P15; 
specifically the proper-motion along Galactic latitude is much smaller in the Subaru-based
measurement. Therefore the Subaru measurement implies a less energetic orbit than that based on 
P15 (see also the next section).
In Table~\ref{tab:tab8} we show the velocity components in cylindrical coordinates $(\Pi, \Theta, W)$ at the location of Draco, 
and in $(U, V)$ along the Sun-Galactic center direction and perpendicular to it.
\begin{table*}
\caption{Velocity components}
\label{tab:tab8}
\begin{tabular}{lrrrrr}
\hline
\multicolumn{1}{c}{Measurement} & \multicolumn{1}{c}{$\Pi$} & \multicolumn{1}{c}{$\Theta$} & \multicolumn{1}{c}{$W$} & \multicolumn{1}{c}{$U$} & \multicolumn{1}{c}{$V$} \\
& \multicolumn{3}{c}{(km~s$^{-1}$)} &  \multicolumn{2}{c}{(km~s$^{-1}$)} \\
\hline
P15 & $27\pm14$ & $89\pm25$ & $-212\pm20$& $-87\pm25$ & $-205\pm14$ \\
Subaru & $-77\pm11$ & $92\pm18$ & $-63\pm17$& $-96\pm18$ & $-309\pm11$ \\
\hline
\end{tabular}
\end{table*}

\subsection{Orbit of Draco}

We calculate Draco's orbit in a 3-component, analytic potential of the Milky Way as 
in e.g., \citet{dgv99}.
Orbital parameters and their uncertainties are derived via Monte Carlo tests in which the initial conditions are
varied based on the uncertainties in the measured
proper motion, radial velocity, and distance of Draco. Upper and lower values of these parameters are representative 
of 1-$\sigma$ errors (i.e., using the $68\%$ interval centered on the median). 

In Table~\ref{tab:tab9} we list the derived orbital period,  apo- and pericentric distances, orbital eccentricity, and 
inclination with respect to the Galactic plane.
\begin{table}
\caption{Orbital parameters}
\label{tab:tab9}
\begin{tabular}{lrrrrr}
\hline
\multicolumn{1}{c}{Measurement} & \multicolumn{1}{c}{$P$} & \multicolumn{1}{c}{$r_a$} & \multicolumn{1}{c}{$r_p$} & \multicolumn{1}{c}{$ecc$} & \multicolumn{1}{c}{$\Psi$} \\
 &  \multicolumn{1}{c}{(Gyr)} & \multicolumn{2}{c}{(kpc)} & & \multicolumn{1}{c}{$(\degr)$} \\
\hline \\
P15 & $2.3^{+0.9}_{-0.5}$ & $136^{+36}_{-23}$  & $64^{+8}_{-9}$ & $0.36^{+0.05}_{-0.05}$ & $70^{+5}_{-6}$ \\ \\
Subaru & $1.4^{+0.2}_{-0.1}$  & $97^{+7}_{-7}$  & $21^{+7}_{-6}$ & $0.66^{+0.06}_{-0.06}$ & $36^{+4}_{-2}$ \\
\hline
\end{tabular}
\end{table}

The Subaru-based orbit is less energetic, of higher eccentricity and lower inclination angle than the $HST$-based one. 
As such, Draco would be
subjected to stronger tidal effects than the $HST$ measurement suggests, 
both due to more frequent pericentric passages and 
to its penetrating farther into the denser regions of the Milky Way.
This is intriguing since Draco has the highest mass-to-light ratio (84~M$_{\odot}/L_{V,\odot}$) among the traditional dwarf-spheroidal
satellites of the Milky Way, with Ursa Minor a close second \citep{mat98,mcc12}.
Draco also displays a break in the light distribution profile \citep{wil04} and a rising velocity 
dispersion profile \citep{wil04,wal12}; this latter feature is seen in only one other dwarf galaxy, Carina (e.g., \citet{mun08}).
These two features are usually taken to indicate some amount of tidal effects (see \citet{mun08} and references therein).
While isophotal stellar density maps of Draco do not show signs of tidal disturbance (S07, \citet{ode01}),
\citet{mun08} argue that this does not preclude tidal disruption, especially when the satellite is at apocenter. 
Our orbit suggests that Draco has just passed apocenter. 
We note that in order to
explain the peculiar velocity-dispersion profile of Draco, \citet{wil04}, have suggested 
an orbit with a pericentric passage of $\sim 20$ kpc, 
with the provision the total mass of the galaxy is modest, $\sim$ few $10^7$ M$_{\odot}$.
Of course, other models that explain the light and velocity dispersion profiles of Draco exist: these however 
advocate large mass-to-light ratios, and an overall large total mass for Draco. Such are the models of 
\citet{mash06} and \citet{jard13}, with mass-to-light ratios of up to 1000~M$_{\odot}/L_{\odot}$ and 
total masses of between $\sim 10^8 - 10^9$ M$_{\odot}$
needed to fit the observed light and velocity dispersion profiles of Draco.  
Our new result for the orbit of Draco suggests that, while the system is dark-matter dominated, perhaps tides do play a role in
lowering the inferred extreme mass-to-light ratio of this system, 
along the lines proposed by \citet{mun08} for Carina and \citet{wil04} for Draco.

\citet{paw13} argue that many MW satellites form a Vast Polar Structure (VPOS), a thin plane 
perpendicular to the MW disk. They also argue that available proper motions indicate that the disk 
is not only a spatial alignment, but also a rotationally supported structure. 
The average direction of the eight most-concentrated orbital poles given by \citet{paw13}
is $(l, b) = (176\degr, -15\degr)$,
with  all eight poles being within  $30\degr$ of this direction.
The $HST$ measurement for Draco places the direction of its orbit pole at 
$(l, b) = (168\degr\pm3\degr, -20\degr\pm4\degr)$. 
Our Subaru-based result gives a direction $(l, b) = (102^{+19}_{-16}, -54^{+5}_{-2})\degr$. 
The large uncertainty in the pole's longitude for our determination stems from the fact that the orbit is 
markedly non-planar compared to that of P15. 
These results indicate that the $HST$ measurement provides an orbit better aligned with the VPOS than ours. 
Nevertheless, our result is not inconsistent with Draco's membership to VPOS \citep[see][figure 1]{paw13}. 
Indeed, the geometry of this structure continues to be refined as more accurate proper motions become available.

\section{Faint, Red, Fast-moving Objects}
Although not primarily directed toward searching for cool dwarf stars, our
proper-motion study is one of the deepest ($i\sim 25$) relatively wide ($\sim 0.2$ square degrees) such study thus far. 
We therefore take the opportunity to
search our proper-motion catalog for fast-moving, red objects as possible brown dwarf candidates. 
Since we are interested in stars with large proper motions where systematics discussed in previous sections have no bearing,
we will use the entire area of our data, and will consider the object classification
from all three epochs' data sets. 
\begin{figure}
\includegraphics[width=\columnwidth,angle=-90]{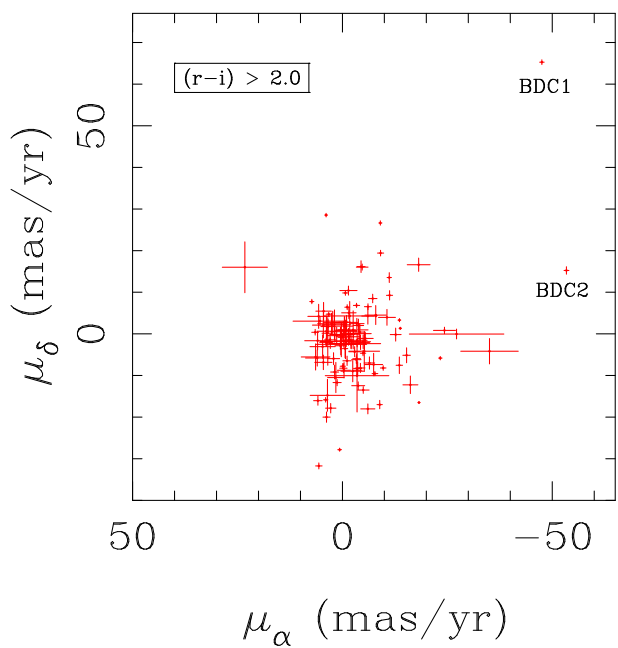}
\caption{Absolute proper motions of stars with $(r-i)_{MegaCam}> 2.0$. 
Two stars are identified as brown-dwarf candidates, having large proper motions
separating them from the remaining population of red stars.}
\label{fig:fig12}
\end{figure}

In Figure~\ref{fig:fig12} we show 
the absolute proper motions of stars with $(r-i)_{MegaCam} > 2.0$. 
The overall population of red stars most likely represents foreground, M-type disk dwarfs.
Two objects stand out in proper-motion space when compared to the motions of the remaining red stars.
They are both well-measured, well-classified stars, with 30 or more frames spanning three epochs.
Their properties are 
summarized in Table~\ref{tab:tab10}, where the photometry and equatorial coordinates (at J2000) are from S07, 
and the absolute proper motions are from this study. 
\begin{table*}
\caption{Properties of Brown Dwarf Candidates}
\label{tab:tab10}
\begin{tabular}{lrrrrrrrrr}
\hline
\multicolumn{1}{c}{Id} & \multicolumn{1}{c}{RA} & \multicolumn{1}{c}{Dec} & \multicolumn{1}{c}{$g_{MegaCam}$} & \multicolumn{1}{c}{$r_{MegaCam}$}& \multicolumn{1}{c}{$i_{MegaCam}$}& \multicolumn{1}{c}{$\mu_{\alpha}$} & \multicolumn{1}{c}{$\mu_{\delta}$} &  \multicolumn{1}{c}{$N_{\mu_{\alpha}}$} &  \multicolumn{1}{c}{$N_{\mu_{\delta}}$} \\
\hline
BDC1 & 17:20:59.7& 57:42:00 & 20.988 & 19.593 & 17.430 &  $-47.53\pm0.53 $ &  $65.27\pm0.57 $ & 55 & 56 \\
BDC2 & 17:19:51.2 & 57:49:39 & 24.955 & 23.474 & 20.809 & $-53.38\pm0.59 $ &  $15.24\pm0.90 $ & 30 & 31 \\
\hline
\end{tabular}
\end{table*}

These objects do not appear in the SIMBAD database, or the Johnston list of brown dwarfs 
database\footnote{http://www.johnstonsarchive.net/astro/browndwarflist.html}. 
While BDC1 is in the ALLWISE source catalogue \citep{mainz11}
with clear detections in bands $W1$ and $W2$, 
BDC2 does not appear in this catalogue. 
For BDC1, the ALLWISE magnitudes of $W1=13.637\pm0.133$ and $W2=13.399\pm0.026$ suggest that the object has a 
spectral type between M6 and L2 
\citep{du12,ga15}.
Using the absolute magnitude calibration from \citet{du12}, we infer a distance of between 95 and 48 pc,
based on these two spectral types, respectively.
Given the high proper motion with respect to stars of similar color, we think it more likely to be closer to
the cooler spectral type, and thus nearer the lower end of that distance range. 
The galactic coordinates imply a geometry in which the line-of-sight velocity translates into rotational velocity 
about the Galaxy, and in which the proper motion along Galactic latitude is mostly projected onto 
the $W$ component. 
If we adopt a heliocentric radial velocity of 0 km~s$^{-1}$ (i.e., the star is moving about the Galaxy much 
like the Sun), then
the vertical velocity is $W = 17$ km~s$^{-1}$ for the lesser distance and $W = 26$ km~s$^{-1}$ for the greater distance.
Both scenarios are thus compatible with membership to the thin disk, although at the greater distance,
one must also consider the old disk, which has a higher velocity dispersion than the younger populations in the thin disk.

The second candidate, BDC2, is somewhat redder and, as such, is a stronger brown-dwarf candidate; its color implying a
temperature cooler than that of a typical L2 dwarf.
It is also relatively faint, at $i = 20.8$. 
Only two or three current brown-dwarf studies, taken in the IR with CFHT and Subaru telescopes,
reach magnitudes fainter that this \citep{kakazu2010,albert2011,casewell2012}.
Thus we see great potential for deep, Subaru-based proper-motion studies
in discovering brown dwarf candidates.

\section{Conclusions}
We have used a large set of deep images taken with Suprime-Cam on the Subaru 8-m telescope 
to measure the absolute proper motion of the Draco dSph galaxy. The data comprise two- and three-epoch determinations
with a time baseline of 4.4 and 7 years. While the exposures reach a limiting magnitude of $i\sim 25$, we limit
the proper-motion study to $i=24.5$.  The area coverage of the entire catalog is 0.2 square degrees; that used for the
determination of the proper motion of Draco is 0.148 square degrees.  Importantly, the images provide
thousands of background galaxies and Draco members with which we perform unprecedentedly detailed tests 
to explore systematic errors in the proper motions.
We obtain an absolute proper motion of high precision, comparable to that of a recent 
2-year baseline $HST$ determination.
Our result disagrees with the $HST$ measurement in $\mu_{\alpha}$,
at the $\sim 6\sigma$ level.
While we discuss
possible sources of error, we cannot identify a cause for this discrepancy.
Because of this difference in motion, the orbit determined here is significantly different from
that based on the $HST$ measurement. Our Subaru-based orbit is less energetic, and thus
more disruptive than the $HST$ one. Our orbit with a pericentric passage of $\sim 20$ kpc suggests 
a scenario in which tides may, in part, account
for the increasing line-of-sight velocity dispersion in the outer regions of Draco \citep{wil04}.

We have also searched for brown-dwarf stars in our catalogue and find two likely candidates, based on their
large proper motions and red colors. 

The full set of proper motions derived in this study is availabe upon request to the first author.

\section*{Acknowledgments}
This work was funded, in part, by the NASA Connecticut Space Grant College Consortium matched with equal funds from 
Southern Connecticut State University.
It makes use of data collected at the Subaru Telescope and obtained from the SMOKA, which is operated by the 
Astronomy Data Center, National Astronomical Observatory of Japan.
We thank the anonymous referee whose suggestions helped improve this manuscript.





\begin{thebibliography}{99}
\bibitem[\protect\citeauthoryear{Albert et al.}{2011}]{albert2011} Albert, L., Artigau, E., Delorme, P., Reyl\'{e}, C., Forveille, T., Delfosse, X., Willott, C. J. 2011, \aj, 141, 203
\bibitem[\protect\citeauthoryear{Armandroff, Olszewski \& Pryor}{1995}]{ar95}Armandroff, T. E., Olszewski, E. W., Pryor, C. 1995, \aj, 110, 2131
\bibitem[\protect\citeauthoryear{Baba et al.}{2002}]{baba02}Baba, H. et al. 2002, ADASS XI, eds. D. A. Bohlender, D. Durand, T. H. Handley, ASP Conference Series, Vol.281, 298
\bibitem[\protect\citeauthoryear{Bertin \& Arnouts}{1996}]{be96}Bertin, E.  Arnouts, S. 1996, \aaps, 117, 39
\bibitem[\protect\citeauthoryear{Bovy et al.}{2012}]{bo12}Bovy, J. et al. 2012, \apj, 759, 131
\bibitem[\protect\citeauthoryear{Carlin et al.}{2012}]{car12}Carlin, J. L., Majewski, S., R., Casetti-Dinescu, D. I., Law, D. R., Girard, T. M.,  Patterson, R. J., 2012, \apj, 744, 25 
\bibitem[\protect\citeauthoryear{Casetti-Dinescu et al.}{2006}]{cd06}Casetti-Dinescu, D. I., Majewsi, S. R., Girard, T. M., Carlin, J. L., van Altena, W. F., Patterson, R. J. Law, D. R., 2006, \aj, 132, 1768
\bibitem[\protect\citeauthoryear{Casewell et al.}{2012}]{casewell2012} Casewell, S. L., Baker, D. E. A., Jameson, R. F., Hodgkin, S. T., Dobbie, P. D., Moraux, E. 2012, \mnras, 425, 3112
\bibitem[\protect\citeauthoryear{Dehnen \& Binney}{1998}]{db98}Dehnen, W., Binney, J. J. 1998, \mnras, 298, 387
\bibitem[\protect\citeauthoryear{Dupuy \& Liu}{2012}]{du12}Dupuy, T. J. Liu, M. C. 2012, \apjs, 201, 19
\bibitem[\protect\citeauthoryear{Dinescu, Girard, \& van Altena}{1999}]{dgv99}Dinescu, D. I., Girard, T. M., van Altena, W. F. 1999, \aj, 117, 1792
\bibitem[\protect\citeauthoryear{Eichhorn}{1988}]{eich88} Eichhorn, H. 1988, in IAU Symp. 133, 177, ed. Suzanne Debarat
\bibitem[\protect\citeauthoryear{Flesch}{2015}]{fle15}Flesch, E. W. 2015, PASA 32, 10
\bibitem[\protect\citeauthoryear{Gagn\'{e} et al.}{2015}]{ga15}Gagn\'{e}, J., LaFreni\`{e}re, D., Doyon, R., Malo, L., Artigau, E. 2015, \apj, 798, 73
\bibitem[\protect\citeauthoryear{Jardel et al.}{2013}]{jard13} Jardel, J. R., Gebhardt, K., Fabricius, M. H., Drory, N., and Williams, M. J. 2013, \apj, 763, 91 
\bibitem[\protect\citeauthoryear{Kakazu et al.}{2010}]{kakazu2010} Kakazu, Y., Hu, E. M., Liu, M. C., Wang, W-H., Wainscoat, R. J., Capak, P. L. 2010, \apj, 723, 184
\bibitem[\protect\citeauthoryear{Kinemuchi et al.}{2008}]{kin08}Kinemuchi, K., Harris, H. C., Smith, H. A., Silbermann, N. A., 
Snyder, L. A., LaCluyz\'{e}, A. P.,  Clark, C. L. 2008, \aj, 136, 1921
\bibitem[\protect\citeauthoryear{Koposov et al.}{2015}]{kop15}Koposov, S. E., Belokurov, V., Torrealba, G., Evans. N. W. 2015, \apj, 805, 130
\bibitem[\protect\citeauthoryear{Lee \& van Altena}{1983}]{leeva83}Lee, J.-F., van Altena, W. F. 1983, \aj, 88, 1683
\bibitem[\protect\citeauthoryear{Lokas, Mamon \& Prada}{2005}]{lok05} Lokas, E. L., Mamon, G. A., and Prada, F. 2005, \mnras, 363, 918
\bibitem[\protect\citeauthoryear{Mainzer et al.}{2011}]{mainz11}Mainzer, A. et al. 2011, \apj, 731, 53
\bibitem[\protect\citeauthoryear{Martin et al.}{2015}]{mar15}Martin, N.F., Nidever, D. L., Besla, G. et al. 2015, \apj, 804, L5
\bibitem[\protect\citeauthoryear{Maschenko, Sills,  \& Couchman}{2006}]{mash06}Maschenko, S., Sills, A., and Couchman, H. M. P., 2006, \apj, 640, 252
\bibitem[\protect\citeauthoryear{Mateo}{1998}]{mat98}Mateo, M. 1998, ARAA, 36, 435
\bibitem[\protect\citeauthoryear{McConnachie}{2012}]{mcc12}McConnachie, A. 2012, \aj, 144, 4
\bibitem[\protect\citeauthoryear{Miyazaki et al.}{2002}]{miy02}Miyazaki, S. et al. 2002, Subaru Prime Focus Camera --- Suprime-Cam, PASJ, 54, 833
\bibitem[\protect\citeauthoryear{Mu\~{n}oz, Majewski \& Johnston}{2008}]{mun08} Mu\~{n}oz, R. R, Majewski, S. R., Johnston, K. V. 2008, \apj, 679, 346
\bibitem[\protect\citeauthoryear{Odenkirchen et al.}{2001}]{ode01}Odenkirchen, M. et al. 2001, \aj, 122, 2538
\bibitem[\protect\citeauthoryear{Ouchi et al.}{2004}]{ou04}Ouchi, M. et al. 2004, \apj, 611, 685
\bibitem[\protect\citeauthoryear{Pawlowski \& Kroupa}{2013}]{paw13} Pawlowski, M. S., Kroupa, P. 2013, \mnras, 435, 2116
\bibitem[\protect\citeauthoryear{Piatek et al.}{2002}]{pia02}Piatek, S., Pryor, C., Armandroff, T. E., Olszewski, E. W. 2002, \aj, 123, 2511
\bibitem[\protect\citeauthoryear{Pryor, Piatek \& Olszewski}{2015}]{pr15}Pryor, C., Piatek, S., Olszewski, E. W. 2015, \aj, 149, 42
\bibitem[\protect\citeauthoryear{Raghavan, et al.}{2010}]{ragh10}Raghavan, D., McAlister, H. A., Henry, T. J., Latham, D. W., Marcy, G. W., 
Mason, B. D., Gies, D. R., White, R. J., and ten Brummelaar, T. A. 2010, \apjs, 190, 1
\bibitem[\protect\citeauthoryear{Robin, et al.}{2003}]{rob03}Robin, A. C., Reyl\'{e}, Derri\`{e}re, S., Picaud, S. 2003, \aap, 409, 523
\bibitem[\protect\citeauthoryear{S\'{e}gall et al.}{2007}]{se07} S\'{e}gall, M, Ibata, R. A., Irwin, M. J., Martin, N. F., Chapman, S. 2007, \mnras, 375, 831
\bibitem[\protect\citeauthoryear{van der Marel}{2015}]{vdm15}van der Marel, R. P. 2015, IAU Symp. 331, 1
\bibitem[\protect\citeauthoryear{Walker}{2012}]{wal12}Walker, M., G., 2013,in ``Planets, Stars and Stellar Systems'', eds. Oswald T. G, and Gilmore, G. Springer Media Dordrecht, vol 5, p. 1039
\bibitem[\protect\citeauthoryear{Wilkinson et al.}{2004}]{wil04}Wilkinson, M. I., Kleyna, J. T., Evans, N. W., Gilmore, G. F., Irwin, M. J.,  Grebel, E. K. 2004, \apj, 611, L21
\bibitem[\protect\citeauthoryear{Yagi et al.}{2002}]{ya02}Yagi, M. et al. \aj, 123, 87
\end{thebibliography}








\bsp	
\label{lastpage}
\end{document}